\documentclass[conference]{IEEEtran}
\usepackage{booktabs} 
\usepackage{tikz}

\newcommand{\question}[1]{\tikz[baseline=(myanchor.base)] \node[circle,fill=.,inner sep=1pt] (myanchor) {\color{-.}\bfseries\footnotesize #1};}

\usepackage{csquotes}
\usepackage{colortbl}
\usepackage{balance}
\usepackage{hyperref}
\usepackage{csquotes}
\usepackage{subcaption}
\usepackage[noend]{algorithm2e}
\usepackage{graphicx}

\usepackage{bbm, dsfont}
\usepackage{amsfonts,amssymb,amsmath}
\newtheorem{example}{\textbf{Example}}
\newcommand{\xhdr}[1]{\vspace{0.75mm}\noindent{{\bf #1}}}

\let\labelindent\relax \usepackage{enumitem}
\def\bitcoin{%
  \leavevmode
  \vtop{\offinterlineskip 
    \setbox0=\hbox{B}%
    \setbox2=\hbox to\wd0{\hfil\hskip-.03em
    \vrule height .3ex width .15ex\hskip .08em
    \vrule height .3ex width .15ex\hfil}
    \vbox{\copy2\box0}\box2}} 
\newcommand{\TDA}[3]{\mathbb{TDA}_{#1}^{#2|#3}}  
\usepackage{color}
\definecolor{shade}{rgb}{0.5,0.8,0.9}
 
\definecolor{Gray}{gray}{0.85}
\newcolumntype{a}{>{\columncolor{Gray}}c}
\newcolumntype{b}{>{\columncolor{white}}c}

\begin{document}
\title{BitcoinHeist: Topological Data Analysis for Ransomware Detection on the Bitcoin Blockchain}


 \author{\IEEEauthorblockN{Cuneyt Gurcan Akcora}
\IEEEauthorblockA{University of Texas at Dallas\\
cuneyt.akcora@utdallas.edu}
\and
\IEEEauthorblockN{Yitao Li}
\IEEEauthorblockA{University of Texas at Dallas\\
yitao.li@utdallas.edu}
\and
\IEEEauthorblockN{Yulia R. Gel}
\IEEEauthorblockA{University of Texas at Dallas\\
ygl@utdallas.edu}
\and
\IEEEauthorblockN{Murat Kantarcioglu}
\IEEEauthorblockA{University of Texas at Dallas\\
muratk@utdallas.edu}}

\maketitle

\begin{abstract}
Proliferation of cryptocurrencies (e.g., Bitcoin) that allow pseudo-anonymous transactions, has made it easier for ransomware developers to demand ransom by encrypting sensitive user data. The recently revealed strikes of ransomware attacks have already resulted in significant economic losses and societal harm across different sectors, ranging from local governments to health care.

Most modern ransomware use Bitcoin for payments. However, although Bitcoin transactions are permanently recorded and publicly available, current approaches for detecting ransomware depend only on a  couple of heuristics and/or tedious information gathering steps (e.g., running ransomware to collect ransomware related Bitcoin addresses). To our knowledge, none of the previous approaches have employed advanced data analytics techniques to automatically detect ransomware related transactions and malicious Bitcoin addresses.

By capitalizing on the recent advances in topological data analysis, we propose an efficient and tractable data analytics framework to automatically detect new malicious addresses in a ransomware family, given only a limited records of previous transactions. Furthermore, our proposed techniques exhibit high utility to detect the emergence of new ransomware families, that is, ransomware with no previous records of transactions. Using the existing known ransomware data sets, we show that our proposed methodology provides significant improvements in precision and recall for ransomware transaction detection, compared to existing heuristic based approaches, and can be utilized to automate ransomware detection.   
\end{abstract}

\section{Introduction}
This decade has been marked with the rise of blockchain based technologies. In its core, blockchain is a distributed public ledger that stores transactions between two parties without requiring a trusted central authority. On a blockchain, two unacquainted parties can create an immutable transaction that is permanently recorded on the ledger to be seen by the public. 
One of the first applications of Blockchain has been the Bitcoin cryptocurrency~\cite{nakamoto2008bitcoin}. Bitcoin's success has ushered an age known as the Blockchain 1.0~\cite{swan2015blockchain}, and currently there exist more than 1000 Blockchain based cryptocurrencies. 

Bitcoin transactions can be created anonymously, and participation in the network does not require identity verification. A payment can be requested by delivering a public Bitcoin address (i.e., a short string) to a sender by using anonymity networks such as Tor~\cite{dingledine2004tor}. This ease of usage and worldwide transaction availability of Bitcoin have been noticed by malicious actors as well. The pseudo-anonymity of cryptocurrencies has attracted the interest of  
a diverse body of criminals, transnational terrorist groups, and illicit users. Cryptocurrency related crime and, more generally, criminal abuse of blockchain technologies are nowadays recognized as the fastest-growing type of cybercrime~\cite{APWG, WSJ2018}.

Among the malicious usage of cryptocurrencies, ransomware payments continue to attract an ever increasing attention.
 Although encrypting files and resources for ransom has a long history, receiving ransom payments securely had never been simple until the emergence of Bitcoin. For example, in 1989 the first documented ransomware AIDS Trojan  demanded payments via international money order or cashier’s check sent to a P.O. box in Panama. Worldwide transactability of Bitcoin protects ransomware operators from revealing their identity to collect the payments. As Paquet-Clouston et al.~\cite{paquet2018ransomware} state, \textquote{The combination of strong and well-implemented cryptographic techniques to take files hostage, the Tor protocol to communicate anonymously, and the use of a cryptocurrency to receive unmediated payments provide altogether a high level of impunity for ransomware attackers}. Starting with CryptoLocker in 2013, the world has seen an explosion of ransomware that uses Bitcoin. Many organizations that have been hit by ransomware are asked to pay significant amounts using Bitcoin. For example, Jackson County, Georgia hit with Ryuk in March, 2019 needed to pay bitcoins equal to \$400,000, since attackers also compromised the backup data~\cite{georgia-ransomware}.
 
Furthermore, using cryptocurrenies for ransomware payments appears to be substantially more prevalent than have been previously realized. 
As noted by Hernendex-Castro et al.~\cite{surveykent}, among the respondents to their survey, \textquote{the prevalence of the Cryptolocker ransomware (3.4\%) seems much higher than expected. The proportion of Cryptolocker victims that claim to have agreed to pay the ransom to recover their files (41\%) seems to be much larger than expected (3\% was conjectured by Symantec, 0.4\% by Dell SecureWorks)}. Hence, understanding the ransomware payments and its overall economical impact is an emerging challenge of critical societal importance.

Although there have been efforts to analyze the cryptocurrency transactions such as Bitcoin using various heuristics (e.g., "co-spending" heuristic that is based on the idea that input addresses to the same transaction must belong to the same person since private keys associated with those accounts are needed for creating valid transactions \cite{meiklejohn2013fistful}) to detect ransomware payments, to our knowledge, none of the previous efforts has leveraged advanced data science tools to detect malicious ransomware payments. 

In this paper, \textit{our goal is to identify Bitcoin addresses that are used to store and trade Bitcoins gained through ransomware activities}. To achieve this goal, we propose a scalable data-driven Bitcoin transaction  analytics framework which is substantially more effective in detecting ransomware payment related addresses, compared to the existing heuristic based approaches.

The key thrust behind our proposed approach is the intrinsic capability to observe the complete blockchain graph and, as a result, \textit{to track and analyze dynamics of the associated blockchain topological and geometrical properties at a multi-resolution level}. A natural question arises: \textit{What does local topological blockchain graph structure tells us on anomalous and malicious patterns?} For example, does a suspiciously repeating occurrence of a certain blockchain payment pattern tend to be associated with ransomware payment behavior?  We show that by using local topological information available on the Bitcoin transaction graph, we can achieve significant improvement in detecting malicious Bitcoin addresses associated with ransomware payments.

Significance of our contributions can be summarized as follows:
\begin{itemize}
   \item To our knowledge, this is the first project that leverages advanced data analytics and machine learning techniques for ransomware related bitcoin address detection.
    
    \item We propose a simple, tractable and computationally efficient framework to extract features related to Bitcoin transactions which exhibit high utility in predicting ransomware related activities.
    
    \item Using the ground truth data collected by various studies, we develop a novel topological data analysis (TDA) based ransomware detection approach that delivers significantly higher precision and recall compared to existing heuristic based procedures.
    
    \item In addition to detecting new addresses associated with a known ransomware family, we show that our new methodology could be used to detect the emergence of new ransomware families.
\end{itemize}

\section{Related Work}
\label{sec:related}

The success of Bitcoin~\cite{nakamoto2008bitcoin} has encouraged hundreds of similar digital coins~\cite{tschorsch2016bitcoin}. The underlying Blockchain technology has been adopted in many use cases and applications.  With this rapidly increasing activity, there have been numerous studies analyzing the blockchain technology from different perspectives.   

The earliest results aimed at tracking the transaction network to locate coins used in illegal activities, such as money laundering and blackmailing \cite{androulaki2013evaluating,ober2013structure}. These findings are known as the taint analysis~\cite{di2015bitconeview}.  

Bitcoin provides pseudo-anonymity; although all transactions are public by nature, user identification is not required to join the network. Mixing schemes~\cite{maxwell2013coinjoin,ruffing2014coinshuffle} exist to hide the flow of coins in the network. Research articles have shown that some Bitcoin payments can be traced~\cite{meiklejohn2013fistful}. As a result, obfuscation efforts~\cite{narayanan2017obfuscation} by malicious users have become increasingly sophisticated. 

In ransomware analysis, Montreal~\cite{paquet2018ransomware}, Princeton~\cite{huang2018tracking} and Padua~\cite{conti2018economic} studies have analyzed networks of cryptocurrency ransomware, and found that hacker behavior can help us identify undisclosed ransomware payments. Datasets of these three studies are publicly available. 

Early studies in ransomware detection use decision rules on amounts and times of known ransomware transactions to locate undisclosed ransomware (CryptoLocker) payments~\cite{liao2016behind}. More recent studies are joint efforts between researchers and blockchain analytics companies; Huang et al. identify shared hacker behavior and use heuristics to identify ransomware payments~\cite{huang2018tracking}. The authors estimate that 20000 victims have made ransomware payments. However, these studies do not extract features and build machine learning models to detect ransomware payments and families.

Feature extraction has been studied for ransomware detection in specific domains. In software code analysis, Cryptolock inspects ransomware programs and their activity for malicious characteristics \cite{scaife2016cryptolock}. In this line of work, studies on ransomware for mobile devices extract software code features to catch malicious programs~\cite{martin2018depth,heldroid}. However, these studies are mainly targeted for detecting ransomware before it can infect a system, and do not consider Bitcoin transactions.

\section{Background and Preliminaries}
\subsection{Ransomware}

 Ransomware is a type of malware that infects a victim's data and resources, and demands ransom to release them. In two main types, ransomware can lock access to resources or encrypt their content. In addition to computer systems, ransomware can also infect IoT and mobile devices~\cite{martin2018depth}.
 
 Ransomware can be delivered via email attachments or web based vulnerabilities. More recently, ransomware have been delivered via mass exploits. For example, CryptoLocker used Gameover ZeuS botnet
to spread through spam emails. Once the ransomware is installed, it communicates with a command and control center. Although earlier ransomware used hard-coded IPs and domain names, newer variants may use anonymity networks, such as TOR, to reach a hidden command and control server. 

In the case of asymmetric encryption, the encryption key is delivered to the victim's machine. In some variants, the encryption key is created on the victim's machine and delivered to the command center. 

Once resources are locked or encrypted, the ransomware  displays a message that asks a certain amount of bitcoins to be sent to a bitcoin address. This amount may depend on the number and size of the encrypted resources. After payment, a decryption tool is delivered to the victim. However, in some cases, such as with WannaCry, the ransomware contained a bug that made it impossible to identify who paid a ransomware amount. 

\subsection{Bitcoin Graph Model}
\label{sec:bitcoingraph}
Largely rooted in the existing network analysis methodology, earlier Bitcoin data analytics techniques approached Bitcoin data by creating a graph that employs only a single type of node.  Such analytic procedures are referred to as {\it transaction and address graph approaches}.

In the {\it transaction graph approach}, addresses are ignored and edges are created among transaction nodes~\cite{fleder2015bitcoin,ron2013quantitative}. Naturally, the {transaction graph} is acyclic and a transaction node cannot have new edges in the future.  

In the {\it address graph approach}~\cite{spagnuolo2014bitiodine}, transactions are ignored and edges are created among address nodes. However, Bitcoin does not store input to output coin flows explicitly --  all inputs are gathered at the transaction, and directed to output addresses at once. As a result, inputs of a transaction must be connected to all its output addresses, which may create large cliques if too many addresses are involved in a transaction.

Single node type approaches do not provide a faithful representation of the Blockchain data. The loss of information about addresses or transactions seem to have an impact in predictive models~\cite{greaves2015using,akcora2018forecasting}. In contrast, we model the Bitcoin graph as a heterogeneous network with two node types: addresses and transactions.  

\xhdr{Bitcoin Graph Model}
We consider a directed weighted graph $\mathcal{G} = (V, E, B)$ created from a set of transactions $TX$ and input and output addresses contained in $TX$. On $\mathcal{G}$, $V$ is a set of nodes, and $E \subseteq V\times V$ is a set of edges. $B=\left\{ \textbf{Address}, \textbf{Transaction} \right\}$ represents the set of node types. For any node $u \in V$, it has a node type $\phi(u) \in B$. For each edge 
$e_{u,v} \in E$ between adjacent nodes $u$ and $v$, we have $\phi(u) \neq \phi(v)$, and either $\phi(u)=\left\{\textbf{Transaction} \right\}$ or $\phi(v)=\left\{\textbf{Transaction} \right\}$. That is, an edge $e \in E$ represents a coin transfer between an address node and a transaction node. This heterogeneous graph model subsumes the homogeneous case (i.e., $\left | B \right |=1$), where only transaction or address nodes are used, and edges link nodes of the same type. Here, we focus on the case where
each address node is linked (i.e., input or output address of a transaction) via a transaction node to another address node. We use $\Gamma_a^i$ and $\Gamma_a^o$ to refer to predecessors (in-neighbors) and successors (out-neighbors) of an address $a$, respectively.

\section{Methodology} 
\label{sec:methodology}

In this paper, we state the following five questions to analyze ransomware behavior on the Bitcoin blockchain:  
\question{1}- What features extracted from the Bitcoin network can be used to detect ransomware behavior?
\question{2}- Does a given ransomware family (e.g., Cryptolocker) show the same behavior on the Bitcoin blockchain over time? 
\question{3}- How similar is the behavior of different ransomware operators on the Bitcoin blockchain?
\question{4}- Can we detect Bitcoin ransom payments that are not reported to law agencies or Blockchain Analytics companies? 
\question{5}- Based on the information about existing ransomware families at a given time, can we detect the emergence of a new ransomware on the Bitcoin blockchain?

To address questions \question{1}-\question{5}, we formulate two primary research problems: i) detecting undisclosed payments to addresses that belong to a known ransomware family and ii) predicting the emergence of a ransomware family unknown to the date. We start by stating the notations used in our problem definitions. Symbols used in this manuscript are given in Tab.~\ref{tab:symbol}.

Let $\{a_u\}_{u\in Z^+}$ be a set of addresses, and let each address $a_u$ be associated with a pair $(\vec{x}_u, y_u)$, where $\vec{x}_u\in \mathcal{R}^D$ is a vector of its features and $y_u$ is its label. That is, depending on a setting, $y_u$ can designate a \textit{white} (i.e., non-ransomware) address or a ransomware address.
Furthermore, we associate timestamp $t_u$ to represent the time when the address $a_u$ first appeared in a blockchain transaction. An address can appear in the blockchain multiple times.
Let $f_1,\ldots,f_n$ be labels of known ransomware families (see Table~\ref{tab:summary}) which have been observed until time point $t$, and let $f_0$ be a label of addresses which are \textbf{not} known to belong to any ransomware family and are assumed to be white addresses.
Before time point $t$, if we observe $l$ addresses $a_1,\ldots, a_l$, then we form their $D \times l$-matrix  of features $X_t=\{\vec{x}_1,\ldots,\vec{x}_l\}$ and a vector of labels $Y_t=\{y_1,\ldots,y_l\}\in \{f_0,f_1,\ldots,f_n\}$.

\noindent We formally define our research problems as follows:

\xhdr{Problem 1 [Existing Family Detection]:} Let ${rs}$ be a known ransomware family of interest. Let $\tilde{Y}_t \subseteq Y_t$ be such that $\forall y_j\in \tilde{Y}_t, y_j\in \{f_0, f_{{rs}}\}$ and $\tilde{X}_t \subseteq X_t$ be the corresponding matrix of features. (If at time point $t$, $\tilde{Y}_t \cap \{f_{{rs}}\}=\emptyset$, increase $t$ such that $\tilde{Y}_t$ contains at least one $f_{{rs}}$.) Now, let $\{a_{l+1}, \ldots, a_{l+z}\}$ be a set of addresses whose set of labels $Y_{t'}=\{y_{l+1}, \ldots, y_{l+z}\}$ is unknown, and let
$X_{t'}=\{\vec{x}_{l+1},\ldots, \vec{x}_{l+z}\}$ be a set of their corresponding observed features. Furthermore, let $t'>t$, and $ t<\min\{t_{a_{l+1}}, \ldots, t_{a_{l+z}}\}$. The problem is to predict all addresses
$a_m\in \{a_{l+1}, \ldots, a_{l+z}\}$ such that $y_m=f_{{rs}}$, using their currently available set of features $X_{t'}$ and previous history $(X_t, Y_t)$.

\xhdr{Problem 2 [New Family Prediction]:} Let ${rs}'$ be a new, yet unobserved ransomware family, and $f_{{rs}'}$ be its label. Let $(X_t, Y_t)$ be a pair of the sets of features and labels, respectively, such that at time point $t$, $\forall y_j\in Y_t, y_j\neq f_{{rs}'}$.  

Now, let $\{a_{l+1}, \ldots, a_{l+z}\}$ be a set of addresses whose set of labels $Y_{t'}=\{y_{l+1}, \ldots, y_{l+z}\}$ is unknown, and let
$X_{t'}=\{\vec{x}_{l+1},\ldots, \vec{x}_{l+z}\}$ be a set of their corresponding observed features. Furthermore, let $t'>t$, and $ t<\min\{t_{a_{l+1}}, \ldots, t_{a_{l+z}}\}$. The problem is to predict all addresses
$a_m\in \{a_{l+1}, \ldots, a_{l+z}\}$ such that $y_m \notin \{f_0, f_1, \ldots, f_n\}$ and $a_m$ is associated with the new ransomware $rs$ (i.e., $y_m=f_{rs'}$), 
using their currently available set of features $X_{t'}$ and previous history $(X_t, Y_t)$.

\subsection{Graph features for classification}
 
On the heterogeneous Bitcoin network, the in-neighbors $\Gamma_n^{i}$ of a transaction $tx_n$ is defined as the set of transactions (not addresses) whose one or more outputs are input to transaction $tx_n$. The out-neighbors of $tx_n$ are denoted as $\Gamma_n^{o}$. 
 
A transaction has inputs and outputs; the sum of output amounts of a transaction $tx_n$ is defined as $\mathcal{A}^{o}(n) =\sum\limits_{a_u \in \Gamma_n^o}{{A}_u^{o}(n)}$, where an output address $a_u$ receives ${A}_u^{o}(n)$ coins.

On the Bitcoin network, an address may appear multiple times with different inputs and outputs. An address $u$ that appears in a transaction at time $t$ can be denoted as $a_u^t$. As most addresses appear only once, for sake of simplicity, we omit $t$ throughout our notations when it is clear from the context.  In order to mine address behavior in time, we divide the Bitcoin network into 24 hour long windows by using the UTC-6 timezone as reference. This window approach serves two purposes. First, the induced 24 hour network allows us to capture how fast a given coin moves in the network. The speed is measured by the number of blocks in a the 24 hour window that contain a transaction involving the coin. In maximum, a coin can appear in ~144 blocks (24 hours, ~6 blocks per hour). Coin speed may disclose certain information on transaction intent. For example, coins that are moved too frequently in the network may be involved in money laundering. Second, the temporal order of transactions within the window helps us distinguish transaction activity from different geolocations. Temporal information of transactions, such as the local time, has been found useful to cluster criminal transactions (see Figure 7 in \cite{huang2018tracking}). 

On the heterogeneous Bitcoin network, in each snapshot we extract the following six features for an address: \textit{income, neighbors, weight, length, count, loop}. 

\begin{figure}
    \centering
    \includegraphics[width=1\linewidth]{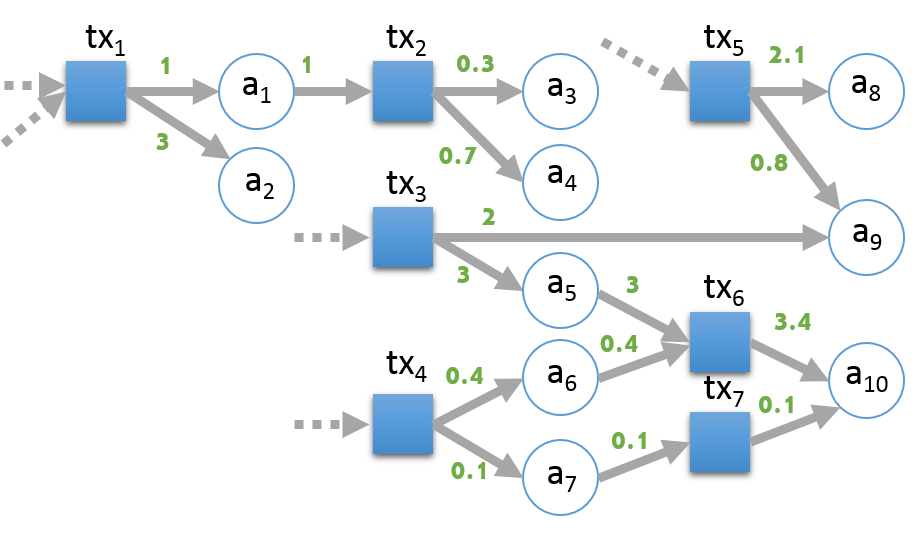}
    \caption{A toy network of 10 addresses and 7 transactions. Dashed edges indicate transaction outputs from earlier windows; $t_1$, $t_3$, $t_4$ and $t_5$ are starter transactions. Coin amounts are shown on edges. Transaction outputs are equal to transaction inputs, i.e., transaction fees are 0.}
    \label{fig:network}
\end{figure}

\xhdr{Income} of an address $u$ is the total amount of coins output to $u$:
$I_u=\sum\limits_{t_n \in \Gamma_u^o}{{A}_u^{o}(n)}$. 

\xhdr{Neighbors} of an address $u$ is the number of transactions which have $u$ as one of its output addresses: $\left | \Gamma^i_u\right |$.

We define the next four address features by using their time ordered position in the defined 24 hour time window. We denote time of a window with the earliest time $t$ of transactions contained in it. For each window, we first locate the set of transactions that do not receive outputs from any earlier transaction within the studied window t, i.e., $\mathbb{TX} = \{ \forall tx_n \in TX , s.t., \Gamma_n^{i}=\{a^{t^0}_1,\ldots,a^{t^n}_z\}, t^0 \leq t^n < t\}$. These transactions consume outputs of transactions that have been generated in previous windows. For simplicity, we refer to a transaction $tx \in \mathbb{TX}$ as a \textbf{starter} transaction.

\xhdr{Weight} of an address $u$, $W_u$, is defined as the sum of fraction of coins that originate from a starter transaction and reach $u$.

\xhdr{Length} of an address $u$, $L_u$, is the number of non-starter transactions on its longest chain, where a chain is defined as an acyclic directed path originating from any starter transaction and ending at address $u$. A length of zero implies that the address is an output address of a starter transaction.

\xhdr{Count} of an address $u$, $C_u$ is the number of starter transactions which are connected to $u$ through a chain, where a chain is defined as an acyclic directed path originating from any starter transaction and ending at address $u$. 

\xhdr{Loop} of an address $u$, $O_u$ is the number of starter transactions which are connected to $u$ with more than one directed path. 

\xhdr{Rationale:} Our graph features are designed to quantify specific transaction patterns. 

Loop is intended to count how many transaction i) split their coins; ii) move these coins in the network by using different paths and finally, and iii) merge them in a single address. Coins at this final address can then be sold and converted to fiat currency (see Figure~7 in~\cite{mcginn2016visualizing} for examples of such patterns). 

Weight quantifies the merge  behavior (i.e., the transaction has more input addresses than output addresses), where coins in multiple addresses are each passed through a succession of merging transactions and accumulated in a final address (see aggregations in Figure~1 of~\cite{huang2018tracking} for a potential application of this pattern). 

Similar to weight, the count feature is designed to quantify the merging pattern. However, the count feature represents information on the number of transactions, whereas the weight feature represents information on the amount (what percent of these transactions' output?) of transactions. 

Length is designed to quantify mixing rounds~\cite{maxwell2013coinjoin} on Bitcoin, where transactions receive and  distribute similar amounts of coins in multiple rounds with newly created addresses to hide the coin origin (see the mixing rounds in Figure 2 of~\cite{ruffing2014coinshuffle}).  

 \begin{example}[Graph Features]
 
Consider the toy network in Figure~\ref{fig:network}. We begin from defining the starter transactions $\mathbb{TX} = \{tx_1,tx_3,tx_4,tx_5\}$ as in this specific window, they receive no coins from an earlier transaction. 

We assign each of the four transactions a weight of 1, regardless of their output amounts $\mathcal{A}^{o}(.)$. Addresses $a_1$ and $a_2$ receive $W_{a_1}=1/|\Gamma^o_{tx_1}|=1/2$ and $W_{a_2}=1/|\Gamma^o_{tx_1}|=1/2$ from transaction $tx_1$, respectively. Address $a_3$ receives $1/2$ of the weight from $a_1$. Hence, $W_{a_3}=(1/2)\times(1/2)$. Consider $a_{10}$ which receives weights from $tx_3$ and $tx_4$. Weight of $tx_3$ flows through $a_5$ and $t_6$, whereas the weight of $tx_4$ flows through $a_7$ and $tx_7$ to reach $a_{10}$. Hence,  $W_{a_{10}}=W_{a_7}+[W_{a_5}+W_{a_6}]=(1/2)+[1/2+1/2]$.

 Length of $a_1$ and $a_2$ is 0, but $L_{a_3}=L_{a_4}=1$, since $a_4$ and $a_5$ can be reached from $tx_1$ through a non starter transaction $tx_2$.  Furthermore, $a_{10}$ can be reached from $tx_3$ or $tx_4$, hence  $L_{a_{10}}=1$. Although $a_8$ and $a_9$ appear later than $a_3$ and $a_4$, lengths of $a_8$ and $a_9$ are shorter, because these addresses are direct outputs of the starter transaction $tx_5$.
 
 The address $a_{10}$ has three chains: i) $tx_3\rightarrow a_5\rightarrow tx_6 \rightarrow a_{10}$,  ii) $tx_4\rightarrow a_6\rightarrow tx_6 \rightarrow a_{10}$ and $tx_4\rightarrow a_7\rightarrow tx_7 \rightarrow a_{10}$, but these chains start from three starters only (i.e., $tx_3$ and $tx_4$). Hence, its count $C_{a_{10}}=2$.
 
 Transaction $tx_4$ has two chains that reach $a_{10}$ through $tx_6$ and $tx_7$, separately. These two chains form a loop from $tx_4$ to $a_{10}$. Hence, $O_{a_{10}}=\left |\{tx_4\}\right|=1$. 
 All other nodes have zero loops.
 \end{example}
 
 \xhdr{Feature Standardization.} In data pre-processing, we use feature standardization to make the values of each feature have zero-mean and unit-variance. We first compute mean and standard deviation for each feature and subtract the mean from each feature value. Then we divide each feature value by its standard deviation.

\subsection{Ransomware address classification and clustering} 
\label{sec:tda}

Before detailing our TDA model, in this section we first outline four classification and clustering models that we employ in this paper. 

\subsection*{Similarity Search:} We use addresses contained in a specific time window $t$, and compute pairwise similarity to known ransomware addresses from the last $l$ days. 

\subsection*{Heuristics} Following two heuristics that are defined by Meiklejohn et al.~\cite{meiklejohn2013fistful} are used in our experimental evaluation.

\xhdr{Co-spending:} \textquote{If two addresses are inputs to the same
transaction, they are controlled by the same user}. 

\xhdr{Transition:} \textquote{If we observe one transaction with addresses A and B as inputs, and another with addresses B and C as inputs, then we conclude that A, B, and C all belonged to the same user}.  

We did not implement the change heuristic proposed in \cite{meiklejohn2013fistful}, because addresses discovered by the change heuristic are not guaranteed to belong to the same user.
 
\subsection*{Unsupervised approaches}

\xhdr{DBSCAN density clustering:} Density-based spatial clustering of applications with noise (DBSCAN) is a density-based non-parametric clustering algorithm. DBSCAN can mark outlier points that lie alone in low-density regions as noise~\cite{ester1996density}. 

\xhdr{Hierarchical clustering:} We use k-means clustering with Forgy algorithm based initial seed selection~\cite{Forgy65}.

\subsection*{Tree based approaches}  

\xhdr{Extreme Gradient Boosting Trees (XGBT):} applies gradient boosting algorithms to decision trees~\cite{chen2016xgboost}.
  
\xhdr{Random forest} is a supervised ensemble of multiple simple decision trees to estimate the dependent variables of the data~\cite{tin1995RF}.

\subsection{Topological Data Analysis Models: TDA Mapper}
\label{sec:TDAMapper}
In this project we introduce the concepts of topological data analysis (TDA) into detection of ransomware patterns on Blockchain. The fundamental idea of TDA is to extract hidden data patterns via systematic analysis of data shapes such as, cycles and flares, quantified at various resolution scales~\cite{Carlsson:2009, lum2013extracting}.

TDA offers the following multi-fold benefits which are particularly useful in the context of ransomware detection. First, TDA assesses data shapes in a coordinate-free manner, which implies that we can systematically compare patterns of data obtained under various data collection frameworks.
Second, TDA analyzes properties of data shapes that are robust to minor data perturbations. Hence, TDA tends to deliver more consistent results on hidden data patterns even under noisy data collection schemes which is a typical scenario in money laundering studies. Finally, TDA offers a low dimensional description of some key properties of the high-dimensional data by using a finite combinatorial object for systematic extraction of data shape patterns. In this paper, we employ the Mapper method of~\cite{singh2007topological, Carlsson:2009} which is a highly customizable TDA tool allowing for a systematic multi-resolution glimpse into organization and functionality behind the underlying data generating process. By complementing more traditional clustering and projection pursuit approaches with a systematic insight on data geometry and topology, Mapper often recovers hidden data patterns that are otherwise inaccessible with conventional data analytic techniques.

The key idea behind Mapper is the following. Let $U$ be a total number of observed addressed and $\{\vec{x}_u\}_{u=1}^U\in \mathcal{R}^D$ be a data cloud of address features. Select a filter function $\xi: \{\vec{x}_u\}_{u=1}^U \to \mathbb{R}$.

Let $I$ be the range of $\xi$, that is, $I = [m, M] \in \mathbb{R}$, where $m = \min_{u} \xi(\vec{x}_u)$ and $M = \max_{u} \xi(\vec{x}_u)$. Now place data into overlapping bins by dividing the range $I$ into a set $S$ of smaller overlapping intervals of uniform length, and let $u_j= \{u : \xi(\vec{x}_u) \in I_j\}$ be addresses corresponding to features in the interval $I_j \in S$. For each $u_j$ perform a single linkage clustering to form clusters $\{u_{jk}\}$.

To find the number of clusters, Mapper analyzes empirical distribution of edge lengths at which each cluster is merged based on the rationale that internal distances (i.e., within a cluster) are expected to be lower than external distances (i.e., in-between clusters) and distributions of internal and external distances are disjoint. Let $\{u_{jk}\}$ be addressed in the $k$-th cluster of the $j$-th interval. View each cluster as a node and draw an edge between two nodes $k$ and $m$ if clusters  $\{u_{jk}\}$ and $\{u_{lm}\}$ contain overlapping addresses, i.e., $\{u_{jk}\}\cap \{u_{lm}\}\neq \emptyset$.

\begin{figure} 
    \centering
    \includegraphics[width=0.9\linewidth]{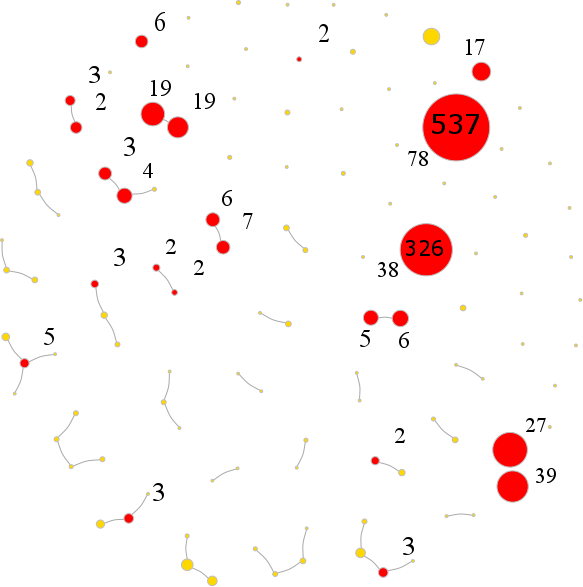}
    \caption{Mapper graph of the Cerber ransomware addresses (2017 day 307), filtered with the length attribute. Around clusters we indicate the number of past ransomware addresses contained in the cluster. We show total order (i.e., address count) of two biggest clusters inside circles (326 and 537). The two clusters with 19 past ransomware addresses (top left) contain 53 and 67 addresses. Clusters without past ransomware are depicted in yellow.}
    \label{fig:tdagraph}
\end{figure}

As a result, Mapper produces a low dimensional representation of the underlying data structure in the form of "cluster tree" graph $\mathcal{CT}$ where each "cluster" is a branch of  some single connected component rather than a disconnected component on its own as in conventional clustering analysis. In Fig.~\ref{fig:tdagraph}, we show an example of the produced Mapper graph. Each node may contain three sets of addresses: past ransomware addresses, past non-ransomware addresses, and addresses of the current time window, whose labels are unknown. If current addresses are contained in clusters that also contain many past known ransomware addresses, we deem these current addresses potential ransomware addresses. 

We filter the TDA mapper graph of using each of our six graph features. As a result, we obtain six filtered graphs $\mathcal{CT}_1,\ldots,\mathcal{CT}_6$ for each time window. Afterwards, 
we assign a suspicion, or risk score to an address $a_u$, as outlined in Algorithm~\ref{alg:tda}. 

\begin{algorithm}
\SetKwData{Left}{left}\SetKwData{This}{this}\SetKwData{Up}{up}
\SetKwFunction{Union}{Union}\SetKwFunction{FindCompress}{FindCompress}
\SetKwInOut{Input}{input}\SetKwInOut{Output}{output}
\Input{A set of networks $\mathcal{CT}_1,\ldots,\mathcal{CT}_D$; filter threshold $q$; inclusion threshold $\epsilon_1$; size threshold $\epsilon_2$; set of past ransomware addresses $RS$; set of past non-ransomware addresses $NRS$ .}
\Output{A set of suspicious addresses.}
$P: Map \leftarrow$ Initialize scores of all $l$ addresses with 0\;
\ForEach{network $\mathcal{CT} \in \{\mathcal{CT}_1,\ldots,\mathcal{CT}_D\}$}{
    \ForEach{cluster $C_c \in \mathcal{CT}$}{
         $A_c \leftarrow $ select all addresses in $C_c$\;
         $V \leftarrow A_c \cap RS$\;
         \If{$\left |V\right | \geq \epsilon_1  \times \left | RS\right | $}{
         \If{$\left |A_c \right | \leq \epsilon_2 \times \left|\mathcal{CT}.V\right|$}{
         \ForEach{$a_u \in A_c\setminus \{RS\cup NRS\}$}{
         $P_u \leftarrow 1+P_u$\;
         }
         }
         }
    }
	}
$q_{t} \leftarrow quantile(P,q)$\;
$S =\{\forall a_u \in P|P_u\geq q_t\}$\; 
return S\;
\caption{TDA filtering with multiple attributes.\label{alg:tda}}
\end{algorithm}

Algorithm~\ref{alg:tda} starts by computing the number of past ransomware addresses in each cluster. If both inclusion and size thresholds,
$\epsilon_1$ and $\epsilon_2$, respectively,
are satisfied, addresses in the cluster have their suspicion scores incremented.  

\xhdr{Parameters.} We use two parameters to control what we learn from mapper clusters: inclusion and size parameters. The inclusion parameter $\epsilon_1$ limits what can be learned when very few ransomware addresses are contained in the cluster. The size threshold $\epsilon_2$ prevents learning when cluster includes too many addresses. Such phenomenon usually happens if a filtering feature does not exhibit a sufficiently discriminating performance during a specific time window, and all addresses are lumped together.  For example, in Figure~\ref{fig:tdagraph} the largest cluster contains 78 past ransomware addresses. Given this finding and the cluster size of 537 addresses, the risk scores of the remaining 459 addresses are to be elevated -- that is, the procedure results in testing of too many addresses as potentially suspicious. We further use a quantile threshold $q$ on addresses, and label addresses suspicious only if they are in the top $1-q$ of all addresses. We emphasize  that by controlling $q$, $\epsilon_1$ and $\epsilon_2$ parameters, we can avoid making predictions when evidence of past ransomware is not sufficiently \textit{strong}. 

We will denote TDA models with the $\TDA{q}{\epsilon_1}{\epsilon_2}$ notation. TDA models may deliver nested results; for example  $\TDA{q}{0.5}{\epsilon_2}$ may return the same set of suspicious addresses as $\TDA{q}{0.7}{\epsilon_2}$ results. For such cases, we prefer the most restrictive model; i.e., the model with the highest $q$, highest $\epsilon_1$ and lowest $\epsilon_2$.  

In summary, the key benefit of the Mapper approach is its capability to recover hidden similarities (i.e., connections) between these "clusters", or groups of addresses, that are typically unavailable with traditional clustering techniques.  

\section{Ransomware Behavior}
\label{sec:experiments}

\xhdr{Dataset.}
We have downloaded and parsed the entire Bitcoin transaction graph from 2009 January to 2018 December. Using a time interval of 24 hours, we extracted daily transactions on the network and formed the Bitcoin graph. We filtered out the network edges that transfer less than $\bitcoin 0.3$, since ransom amounts are rarely below this threshold. As mentioned in Sec.~\ref{sec:elimination}, in final verification we used the full graph to discover smaller amounts that we may have missed due to this filtering. 

  \begin{figure}
       \centering
       \includegraphics[width=0.9\linewidth]{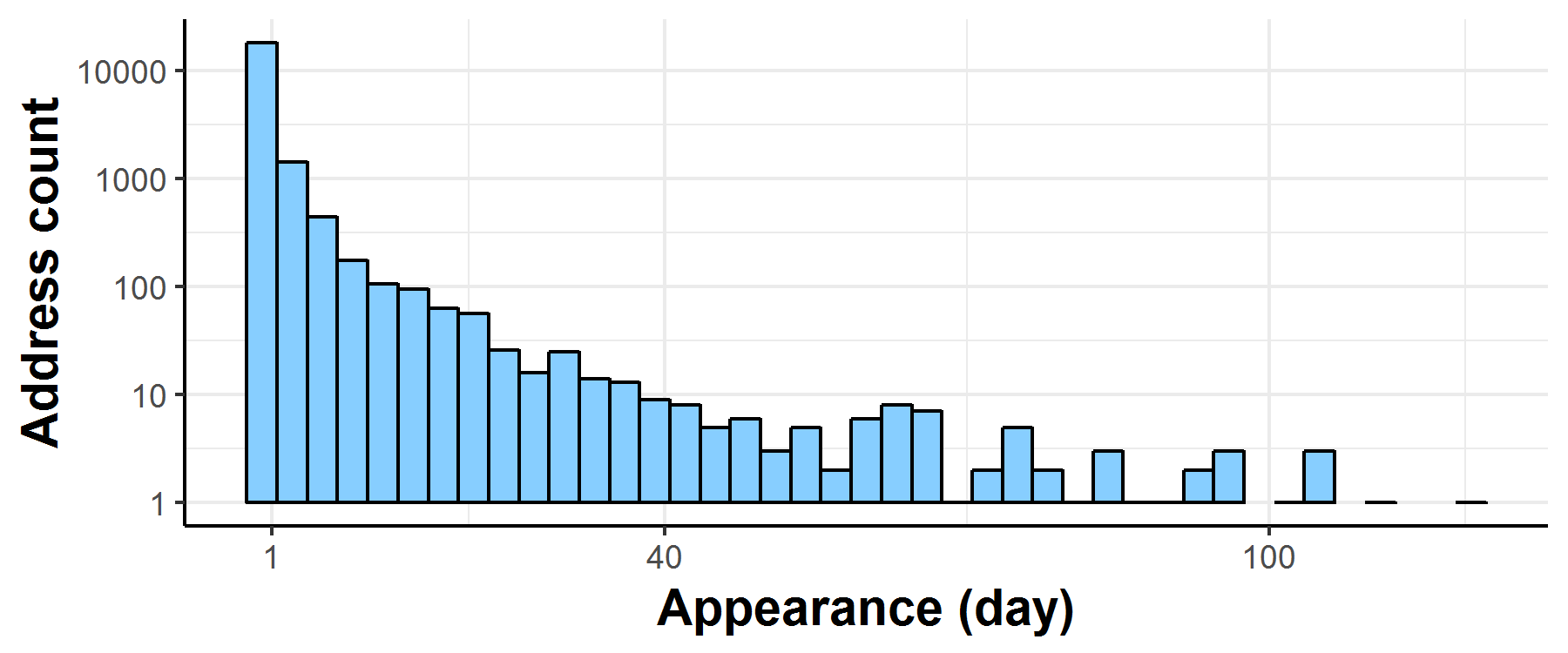}
       \caption{Address recurrence in days. 11 addresses have more than 120 appearances (omitted here).}
       \label{fig:recur}
   \end{figure}

Our ransomware dataset is a union of datasets from three widely adopted studies:    
Montreal~\cite{paquet2018ransomware}, Princeton~\cite{huang2018tracking} and  Padua~\cite{conti2018economic}. The combined dataset contains 24,486 addresses from 27 ransomware families. From these addresses, we extract the features listed in Tab.~\ref{tab:summary}. The column \textit{N} shows the number of times addresses from a given family appear in the Bitcoin blockchain. The \textit{unique} column gives the number of unique addresses for each family. In 24 ransomware families, at least one address appears in more than one 24-hour time window.  Fig.~\ref{fig:recur} shows a histogram of appearances. CryptoLocker  has  13 addresses that appear more than 100 times each. The CryptoLocker address {\tiny 1LXrSb67EaH1LGc6d6kWHq8rgv4ZBQAcpU} appears for a maximum of 420 times.  Four addresses have conflicting ransomware labels between Montreal and Padua datasets. APT  (Montreal) and Jigsaw (Padua) ransomware families have two and one \textit{multisig} addresses, respectively. All other addresses are ordinary addresses that start with '1'.

\subsection{A summary on ransomware features}

\begin{table}[htbp]
  \centering   
  \caption{Most common 6 feature value combinations (i.e. patterns) in ransomware addresses. Address column gives  the number of addresses with the pattern. Rank of a pattern is computed over all (ransomware or not) addresses in the blockchain. Difference in ranks of ransomware and non-ransomware patterns is found statistically significant by two-sample Wilcoxon rank sum test~\cite{bauer1972constructing},  $p$-value $< 2.2\times 10^{-16}$. }
  \setlength\tabcolsep{0.9pt}
  {\small
    \begin{tabular}{ccccclccr}
     Length&Weigth&Neighbor&Count&Loop&{Income}&\#Address&Rank\\
    \midrule
0&0.5&2&1&0&\bitcoin 1 &327&1\\
0&0.5&2&1&0&\bitcoin 1.2&250&113\\
0&1.0&2&1&0&\bitcoin 1  &189&4\\
0&1.0&1&1&0&\bitcoin 0.5&178&9\\
0&0.5&2&1&0&\bitcoin 0.8&160&116\\
0&1.0&1&1&0&\bitcoin 1  &146&3\\
0&1.0&2&1&0&\bitcoin 1.2&127&121\\
0&0.5&2&1&0&\bitcoin 1.25&119&327\\
0&0.5&1&1&0&\bitcoin 0.5&118&6\\
0&1.0&1&1&0&\bitcoin 2  &117&18
    \end{tabular}%
    }
  \label{tab:payments}%
\end{table}%

We compute six feature values of 48,168 ransomware address appearances. We find 30K unique feature values (i.e., unique rows of data), which we term \textbf{patterns}. Table~\ref{tab:payments} lists the top 10 most frequent patterns in ransomware address features. Here, 54 of the top 100 most frequent patterns of all addresses are also found in the top 100 of ransomware patterns. The \textit{Rank} column shows the rank of a pattern in all transactions. We find that four of  the top-10 ransomware address patterns do not appear in the top-100 patterns of all addresses. Furthermore, length=0 implies that ransomware addresses receive payments from starter transactions; in the given window, such addresses have no past history. Finally, weight 0.5 implies that the starter transaction pays the ransomware address, and uses a change address. We also find that a ransomware address typically receives all coins of the transaction, i.e., weight=1.0.  

\begin{figure*}
\begin{subfigure}{.22\textwidth}
  \centering
  \includegraphics[width=1\linewidth]{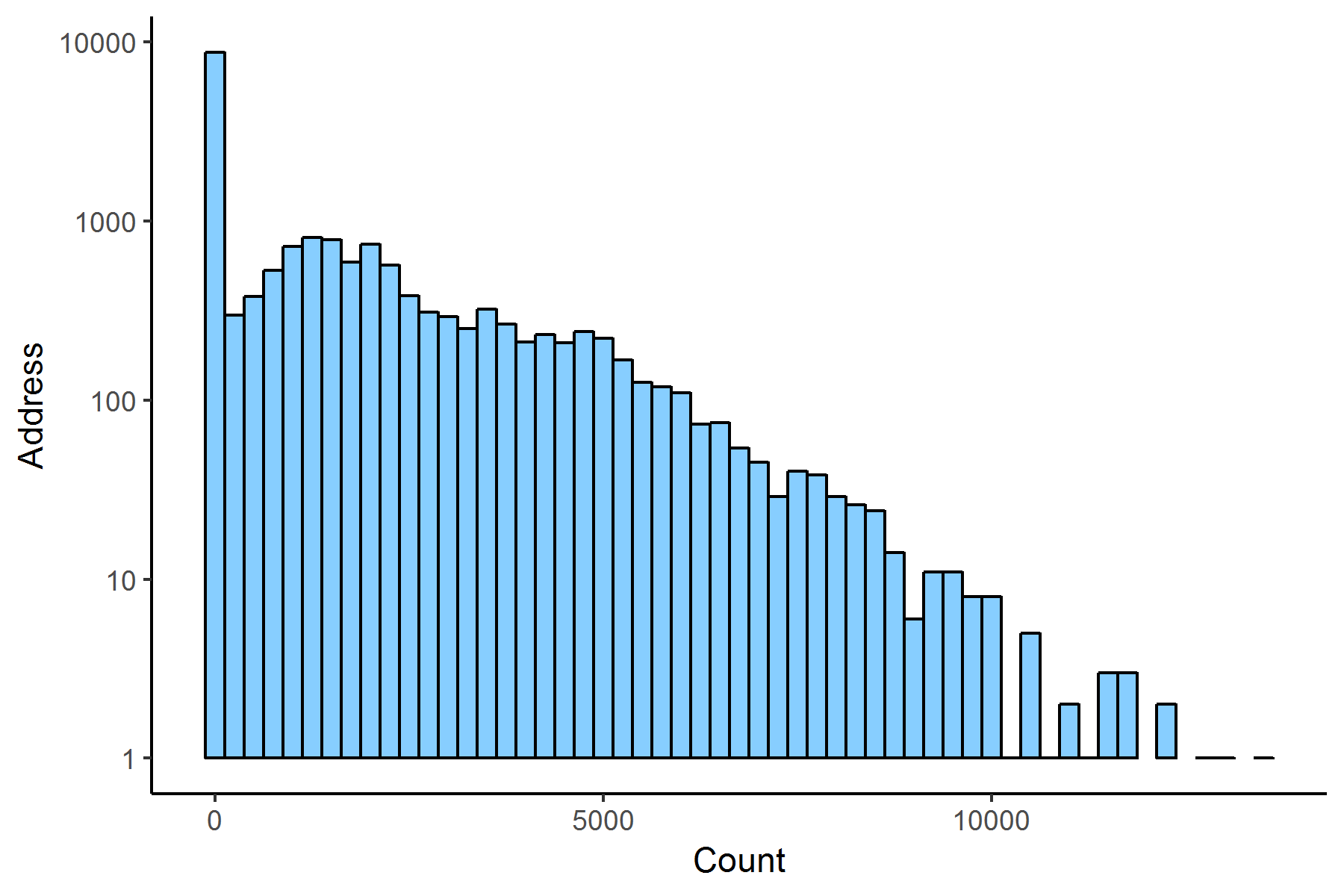}
  \caption{Count distribution in ransomware addresses.}
\end{subfigure}%
\begin{subfigure}{.22\textwidth}
  \centering
  \includegraphics[width=1\linewidth]{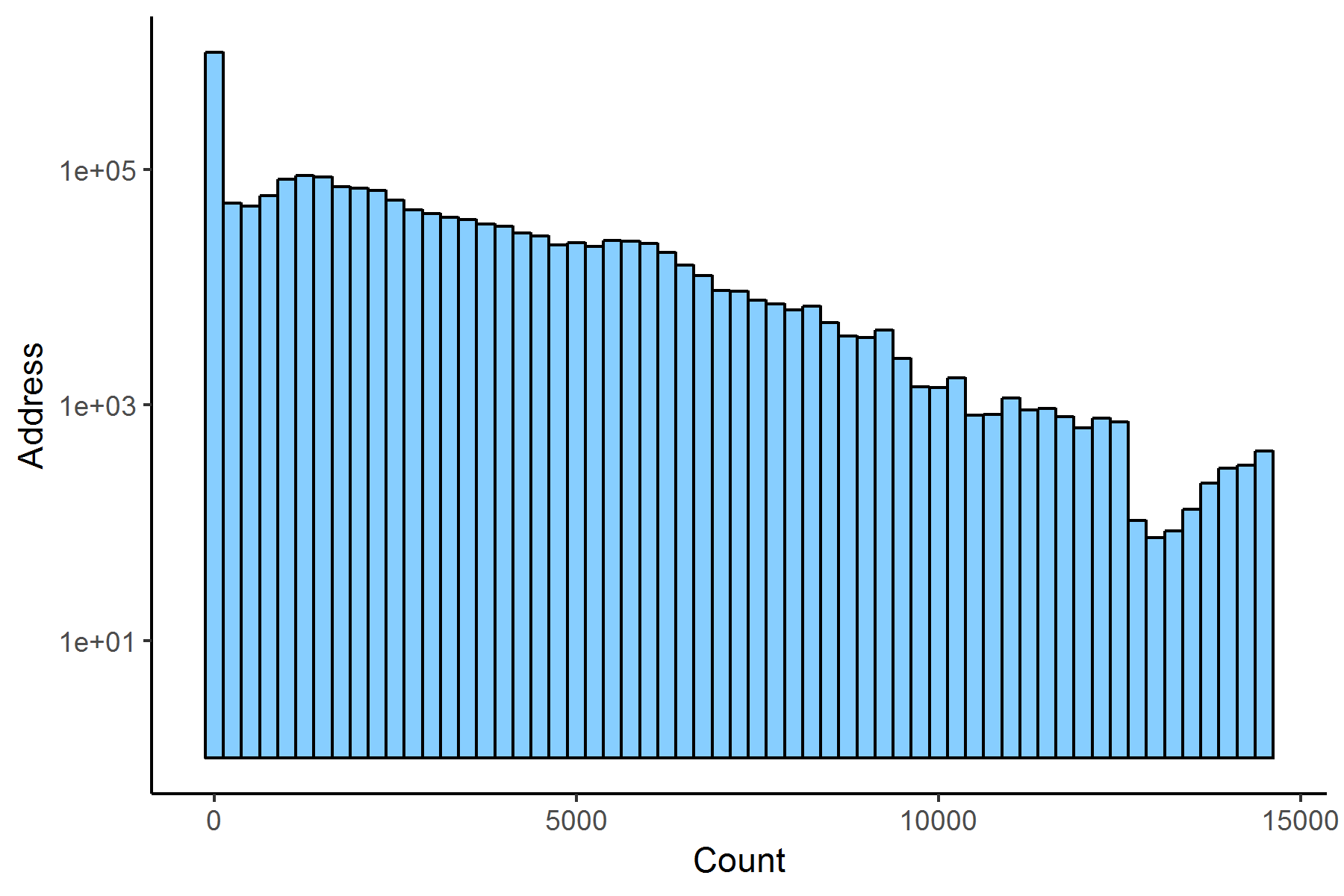}
  \caption{Count distribution in non-ransomware addresses.}
\end{subfigure}
\begin{subfigure}{.22\textwidth}
  \centering
  \includegraphics[width=1\linewidth]{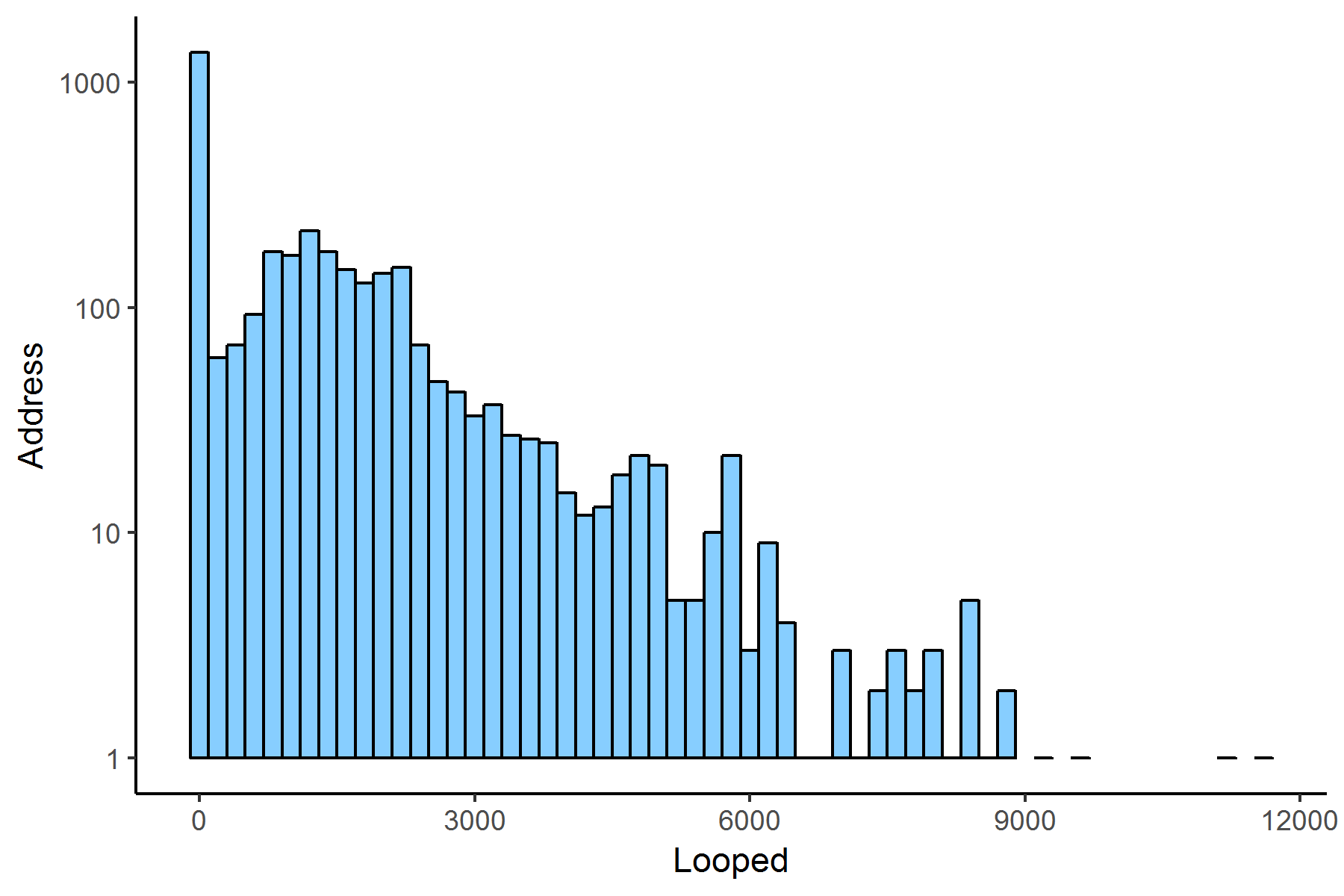}
  \caption{Loop distribution in ransomware addresses.}
\end{subfigure}%
\begin{subfigure}{.22\textwidth}
  \centering
  \includegraphics[width=1\linewidth]{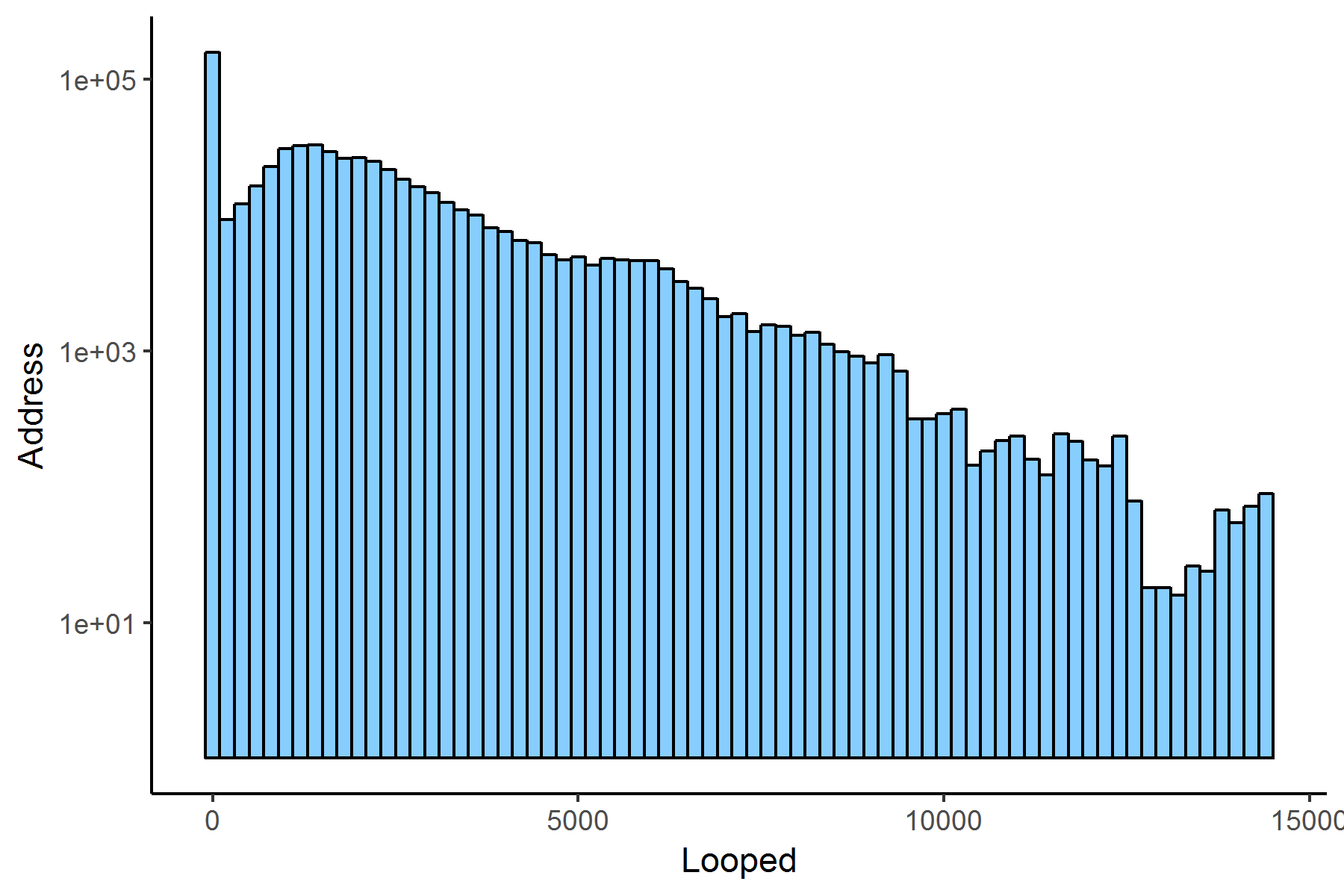}
  \caption{Loop distribution in non-ransomware addresses.}
\end{subfigure}
\caption{Distributions of feature values for ransomware and non-ransomware addresses. Ransomware addresses are skewed for both features. The skew also exists in weight and length features (not shown here).}
\label{fig:features}
\end{figure*}
 
Table~\ref{tab:summary} shows mean values of the graph features for ransomware addresses. CryptoLocker, CryptXXX, Locky, Cerber, DMALockerv3 and CryptoWall have more than 100 unique addresses. The Razy ransomware exhibits an outlier behavior in five features. In weight, only DMALocker has a value $>1$. WannaCry family is an outlier in both length and loop features. Figure~\ref{fig:features} depicts distributions of two features, count and loop. 
When compared to non-ransomware addresses, ransomware addresses exhibit more profound right skewness in distributions of feature values.

\subsection{Feature Relevance for ransomware behavior}
\label{sec:tsne}

 \begin{figure*}[ht!]
 \centering
\begin{subfigure}{.33\textwidth}
 \includegraphics[width=1.0\linewidth]{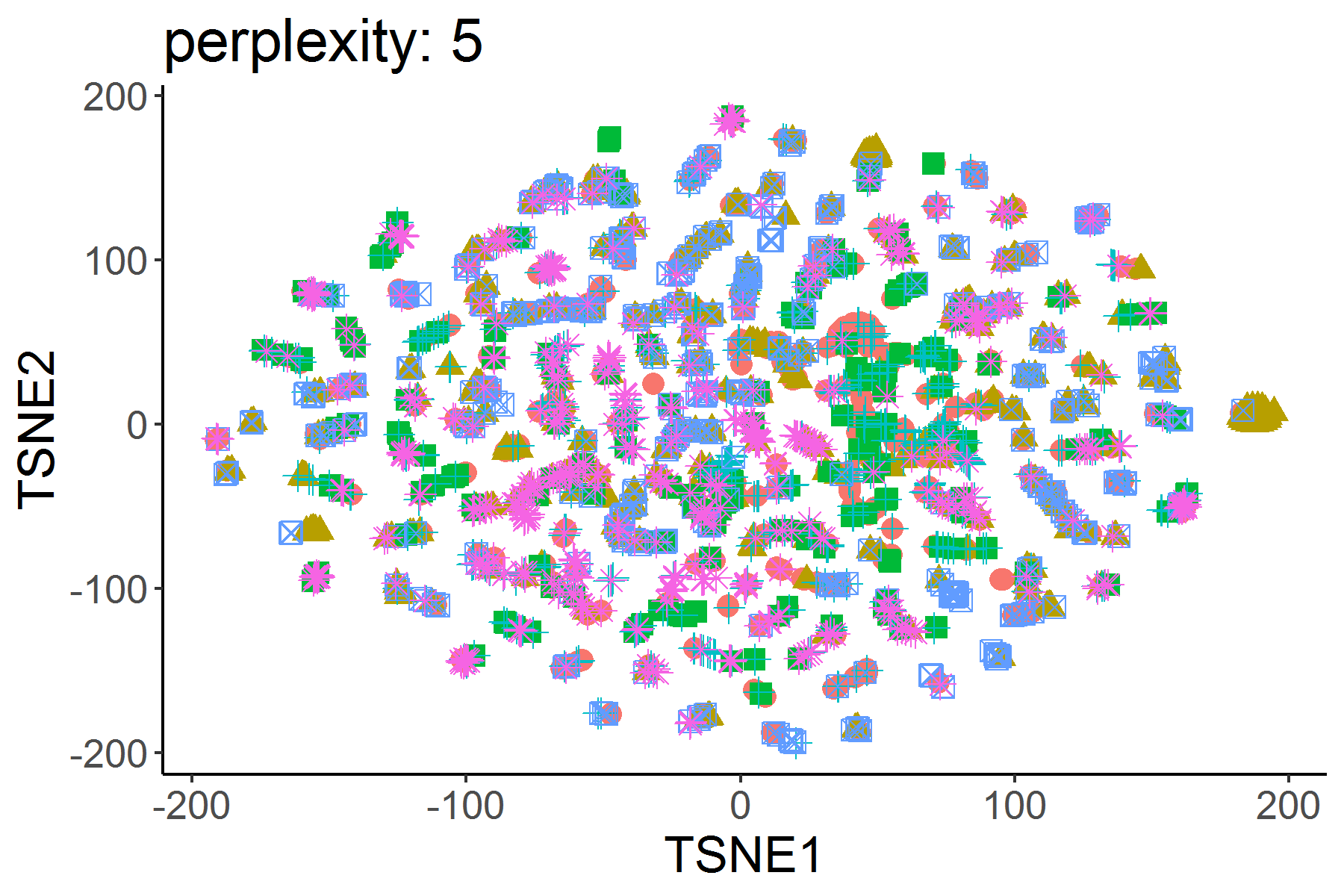}
\end{subfigure}%
~
\begin{subfigure}{.33\textwidth}
   \includegraphics[width=1.0\linewidth]{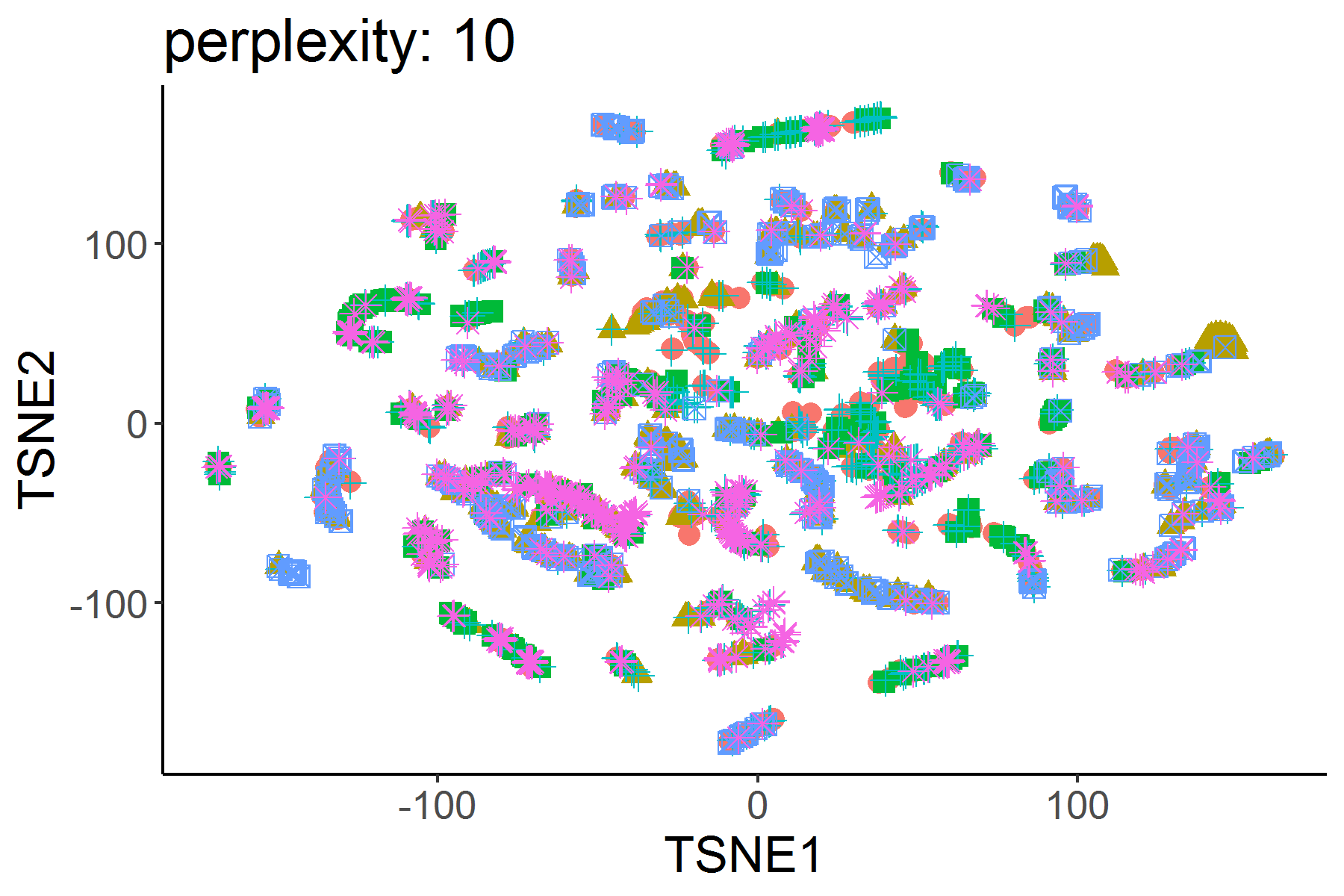}
\end{subfigure}%
~
\begin{subfigure}{.33\textwidth}
   \includegraphics[width=1.0\linewidth]{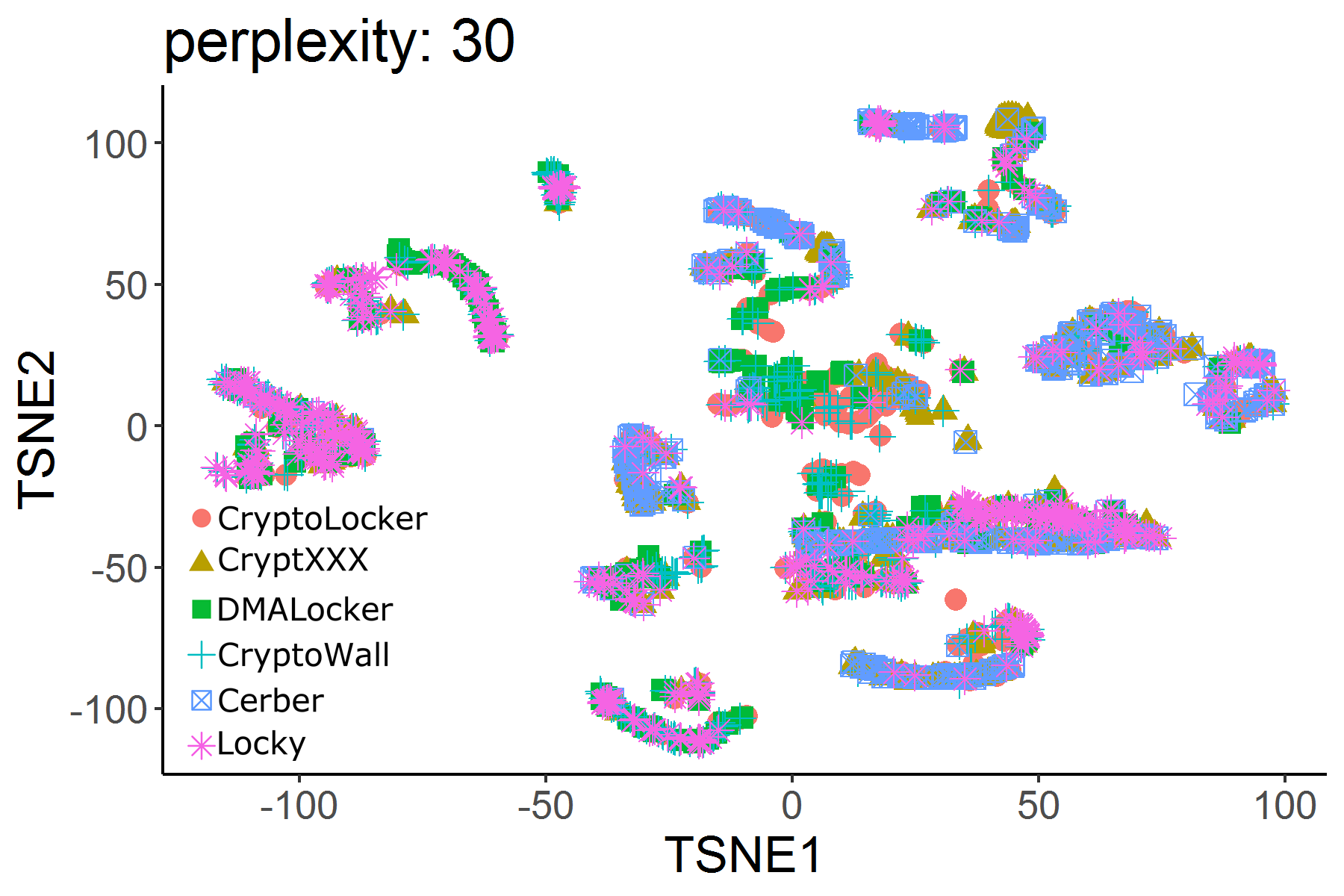}
\end{subfigure}
\caption{Two dimensional t-Stochastic Neighbor embeddings for addresses from six ransomware families. The perplexity can be interpreted as a measure of the effective number of neighbors to be considered for each point.}
\label{fig:tsne}
\end{figure*}

We compute feature values for each appearance of ransomware addresses, and visualize the six-dimensional feature data in Fig.~\ref{fig:tsne} for the six largest families. Each data point is given a location in a two-dimensional map by the t-stochastic neighborhood embedding~\cite{maaten2008visualizing}. We show results for three perplexity values; perplexity   \textquote{can be interpreted as a smooth measure of the effective number of neighbors}~\cite{maaten2008visualizing}. 
Distances and sizes may not be meaningful in TSNE, however, groups of data points exhibit interesting patterns. Each family can be seen to have small groups of co-clustering addresses, implying that a few addresses have similar features. As perplexity increases, we find that addresses tend to cluster together, with DMALocker and CryptXXX addresses appearing at the center.  CryptoWall and CryptoLocker addresses appear together in many groups. The most important insight to be gained from TSNE results is that addresses from a ransomware family do not all exhibit the same behavior -- each family have a multitude of patterns that repeat across addresses.

\subsection{Ransomware behavior similarity}
\label{sec:cluster}

\begin{figure}
       \centering
       \includegraphics[width=0.9\linewidth]{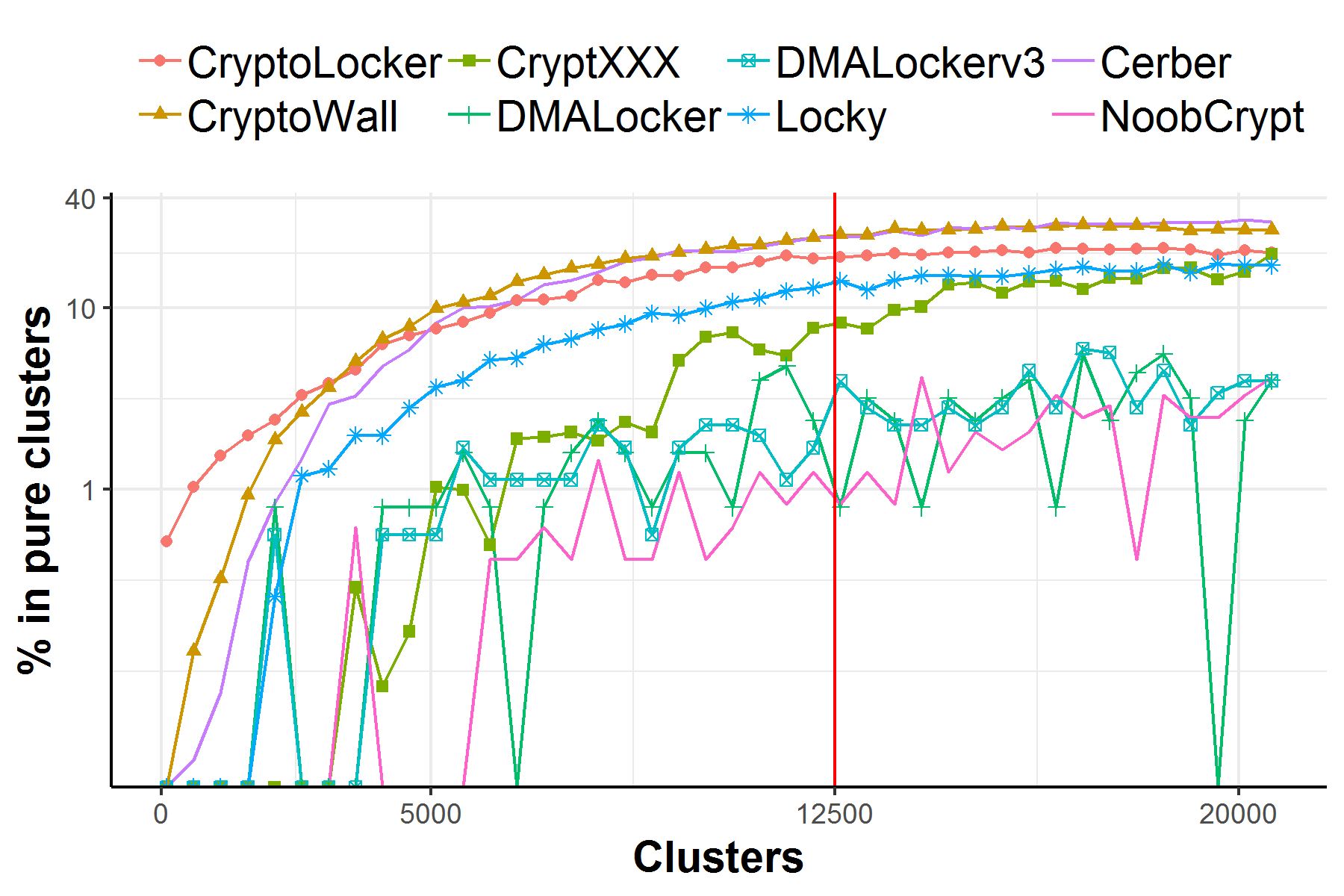}
       \caption{Clustering ransomware addresses by using six features.}
       \label{fig:c1clusters}
   \end{figure}

 We use ransomware labels as external ground truth, and cluster ransomware addresses in up to 20,000 clusters. We compute cluster purity for each family, which we define as the percentage of addresses from a family appearing in clusters where every address in the cluster belongs to the same ransomware family. In computing purity, we only consider clusters that contain more than one data point. Fig.~\ref{fig:c1clusters} depicts purity for the eight biggest ransomware families by address count; Cerber CryptoWall and CryptoLocker reach almost 40\% purity. As in Fig.~\ref{fig:c1clusters}, we find an optimal number of $k=12500$ clusters.  
 
Next, we analyze the resulting clusters for relationships between ransomware families. In Tab.~\ref{tab:sharedclusters}, we show the number of clusters that are shared by ransomware pairs.  CryptoWall addresses appear in 1152 pure clusters (all addresses belong to CryptoWall). CryptoLocker and CryptoWall addresses appear in  2587 impure clusters. In Tab.~\ref{tab:sharedaddresses}, we further look into these shared clusters and count how many addresses are co-clustered in families. We see that 5122 CryptoLocker addresses (1st row, 2nd column) are clustered with 4826 CryptoWall addresses (2nd row, 1st column). From Tab.~\ref{tab:sharedaddresses}, we again see that CryptXXX more often co-clusters with CryptoLocker. The case of CryptoLocker-CryptoWall is intriguing; CryptoLocker addresses cluster more often with CryptoWall addresses than with other CryptoLocker addresses. 

In August 2014, the CryptoLocker ransomware was taken down by Operation Tovar: a consortium constituting FBI, Interpol, security software vendors and several universities. Two months earlier, the CryptoWall ransomware had been  discovered. CryptoLocker spread through email attachments, whereas in addition to attachments, CryptoWall uses kits hosted on compromised websites or malicious ads. CryptXXX appeared in April 19, 2016.  

Clustering uses address features that encode transaction behaviors of two set of users: initial ransom payers, and  hackers that move paid ransom amounts on the network. As initial payers are unknown to each other, we do not expect them to exhibit similar behavior. We believe that the similarity implied by the clustering results can only be due to shared hacker behavior. As such, CryptoLocker, CryptoWall and CryptXXX may be created and run by the same hackers. As a supporting evidence, for CryptXXX, the security company Symantec warns that \textquote{Definitions \textit{(of the detection tool)} prior to September, 2016 may detect this threat as Trojan.CryptoLocker.AN})~\cite{symantecwall} (We could not find such a note for other ransomware families).

\begin{table}[htbp]
  \centering
  \caption{Number of shared clusters between ransomware families ($k=12500$).  At least one address of the ransomware in row is co-clustered with at least one address of the ransomware in columns. If more than two families exist in a cluster, the cluster is counted for each pair.}
  \setlength\tabcolsep{1.4pt}
  \footnotesize
    \begin{tabular}{rcccccccc}
          & {Crypto} & {Crypto} & &  & {Crypt} & {Noob} & {DMA}  \\
           & {Wall} & {Locker} & {Cerber} & {Locky} & {XXX} & {Crypt} & {Locker}  \\
           \hline 
    {CryptoWall} & 1152  & 2587     & 976   & 1403  & 374   & 304   & 129 \\
    {CryptoLocker} &   & 690  & 1220  & 841   & 379   & 142   &  99 \\
    {Cerber} &       &       & 647  & 493   & 562   & 18    &  20 \\
    {Locky} &       &       &       & 296   & 266   & 87    &  55 \\
    {CryptXXX} &       &       &       &       & 49   & 24    &  5 \\
    {NoobCrypt} &       &       &       &       &       & 6    &  28 \\
     {DMALocker} &       &       &       &       &       &      & 2
    \end{tabular}%
  \label{tab:sharedclusters}%
\end{table}%

\begin{table}[htbp]
  \centering
  \caption{Number of co-clustering addresses from ransomware families ($k=12500$).}
  \setlength\tabcolsep{1.4pt}
  \footnotesize
    \begin{tabular}{rcccccccc}
          & {Crypto} & {Crypto} & &  & {Crypt} & {Noob} & {DMA}  \\
           & {Wall} & {Locker} & {Cerber} & {Locky} & {XXX} & {Crypt} & {Locker}  \\
           \hline 
    {CryptoWall} & 3145  & 5122     & 1737   & 3015  & 624   & 702   & 321 \\
    {CryptoLocker} & 4826  & 1766  & 2380  & 1843   & 618   & 238   &  179 \\
    {Cerber} &   3415    &   2285    & 1885  & 9763   & 1696   & 22    &  37 \\
    {Locky} &   4077    &   3023   &    1512   & 832   & 460   & 405    &  384 \\
    {CryptXXX} &  1024     &   1077    &  1467     &    355   & 155   & 52    &  5 \\
    {NoobCrypt} &   344    &   164    &   18    &    95   &   25    & 12    &  37 \\
     {DMALocker} &   140    &   107    &   21    &    61   &   6    &   33   & 4 
    \end{tabular}%
  \label{tab:sharedaddresses}%
\end{table}%

\subsection{Address features over time}

In our detection and prediction problems, we assume that ransomware addresses exhibit similar feature patterns in time, and we can learn to identify these patterns. However, the TSNE results in Sec.~\ref{sec:tsne} show that a global behavior may not exist. In fact, when we compute ransomware patterns, Tab.~\ref{tab:repeating} shows that only 10 families have feature patterns that are repeated in time (see Tab.~\ref{tab:payments} patterns). For example,  Cerber has 3491 addresses use 564 unique patterns more than once. Some of these pattern repeating addresses appear in hundreds of time windows: the address {\tiny 1LXrSb67EaH1LGc6d6kWHq8rgv4ZBQAcpU} appears in 420 windows.  We cluster it in 66 clusters multiple times (as given in Sec.~\ref{sec:cluster}); in one cluster it appears six times. Overall, 100 addresses from CryptoLocker, CryptoWall, NoobCrypt, DMALockerv3, GlobeImposter, DMALocker and CryptXXX families have repeating patterns in time. These results offer evidence that although we cannot talk of a general behavior, small groups of addresses from at least 10 families can be used to discover other ransom addresses.
 
\begin{table}[htbp]
  \centering
  \caption{Repeating patterns in ransomware transactions.}
    \begin{tabular}{lrr}
    Ransomware & {Unique pattern} & {\# addresses}\\
    \midrule
    Cerber & 564   & 3491 \\
    Locky & 403   & 2704 \\
    CryptoWall & 389   & 1406 \\
    CryptoLocker & 248   & 1201 \\
    CryptXXX & 141   & 1018 \\
    DMALockerv3 & 19    & 56 \\
    Globev3 & 3     & 7 \\
    DMALocker & 3     & 6 \\
    GlobeImposter & 1     & 2 \\
    NoobCrypt & 1     & 2 
    \end{tabular}%
  \label{tab:repeating}
\end{table}%
 
\section{Ransomware detection and prediction}

In this section we first discuss two factors that we account for in training our models. 

\xhdr{Class imbalance.} Compared to 24,486 known ransomware addresses from 2013 to 2018, there exist 200K daily transactions on average on the Bitcoin network, where each transaction contains an average of ~5 input and output addresses.  
As a result, we observe a
large number of $f_0$ (i.e, non-ransomware) addresses, leading to a class imbalance problem in classification. To reduce class imbalance, we train on a limited number of addresses; that is, we employ $N \in\{300, 600, 1000\}$ samples of $f_0$ and $f_1,\ldots,f_n$ (i.e., ransomware) addresses each. Our choice is due to the fact that on most days we observe much fewer than 100 unique ransomware addresses. 

\xhdr{Temporal variation.} Ransomware families appear and disappear in certain time periods. For example, the CryptoLocker ransomware was active until 2014 September. Although its clones, such as CryptoWall, appeared afterwards, using CryptoLocker addresses in training for later days may increase false positives. To mitigate these effects, we employ a sliding window approach, and use data from last $l \in \{30,60,90,120,240\}$ windows to train our models. 

\xhdr{Metrics.} For each model and ransomware family, we compute precision and recall by using overall sums of TN, FN, FP and TP values across multiple time windows. Precision and recall are computed as $TP/(TP+FP)$ and $TP/(TP+FN)$, respectively. As we observe considerably more non-ransomware addresses, we identify the positive likelihood ratio ($PLR=TP/FP$) as an important metric, since PLR quantifies the effort needed for analyzing the blockchain by using model prediction results.  For example, a $PLR$ of 1 implies that for every TP address, the analyst has to manually analyze one FP address, unnecessarily. 

\xhdr{Parameters.} In all models, we report the optimal parameters that maximize F1 scores in predictions as follows: In DBSCAN, we experimented with $\epsilon=0.05,\ldots,1$ values. Random Forest uses ntree=500 and mtry=$\left |X_t\right|/3$. XGBoost uses the gbtree booster and nrounds = 25. For TDA computations, we use the TDAMapper RStats package (\url{https://github.com/paultpearson/TDAmapper}) with parameters overlap=40, interval = 80. 
\subsection{Existing family detection} 

We outline our experimental settings for the detection problem as follows.

Given features and known labels of past addresses $X_t, Y_t$ at time $t$ and features of addresses $X_{t^\prime}$ at time $t^\prime >t$, we train for existing ransomware family detection as follows:
\begin{enumerate} [leftmargin=0.1cm,itemindent=.5cm,labelwidth=\itemindent,labelsep=0cm,align=left]

    \item Select a ransomware family $rs$ whose new addresses will be detected at time $t^\prime$.
    \item For $t< t^\prime$, use a training length $l$, and create a dataset $X_t$ which holds features and labels of addresses observed between times $t-l$ and $t$. 
    \item Create an $f_0$ sample of size N from $X^0_{\left[t-l,t\right]} \subseteq X_{\left[t-l,t\right]}$ without replacement where $\forall x_u \in X^0_{\left[t-l,t\right]} , y_u=f_0$ and $N=\left |X^0_{\left[t-l,t\right]}\right|$. 
    \item \label{l:1} Create a \textit{ransomware} sample of size N from $X^{rs}_{\left[t-l,t\right]} \subseteq X_{\left[t-l,t\right]}$ without replacement where $\forall x_u \in X^a_{\left[t-l,t\right]} ,
    \textcolor{red}{y_u=f_{rs}}$ and $N \leq \left |X^{rs}_{\left[t-l,t\right]}\right|$. 
    \item Using the ground truth data at $t^\prime$, find all ransomware addresses for $t^\prime$:  $X^{rs}_{t^\prime}$.
    \item Using the ground truth data at $t^\prime$, take a sample of $M=1000$ white (i.e., $f_0$) addresses without replacement: $X^0_{t^\prime}$.
    \item \label{l:2} Remove past known addresses from $X^{rs}_{t^\prime}$, i.e., $X^{rs}_{t^\prime}\leftarrow X^{rs}_{t^\prime} \setminus X^{rs}_{\left[t-l,t\right]}$.
    \item Use features $\{ X^0_{\left[t-l,t\right]} \cup X^{rs}_{\left[t-l,t\right]} \}$ and labels $\{ Y^0_{\left[t-l,t\right]} \cup Y^{rs}_{\left[t-l,t\right]} \}$ as the training data, and classify $\{ X^0_{t^\prime} \cup X^{rs}_{t^\prime} \}$.
\end{enumerate}

\noindent We emphasize four aspects of existing family detection: i) [Step~\ref{l:2}]: from the test dataset we remove appearances of addresses that have appeared in the past (i.e., $t<t^\prime$), because their labels are already known. ii) [Step~\ref{l:1}]: If an address appears in multiple windows, its each appearance will have (potentially) different features in  $X_{\left[t-l,t\right]}$ with the same $rs$ label. iii) [Step~\ref{l:2}]: On many days, we will not have N past $rs$ addresses to train from. iv) Most importantly, \textit{we learn a model for each ransomware family}. In our experiments, we show that these models do not share the same characteristics. 

\xhdr{Heuristics.} By employing the co-spending and transition heuristics with all past history (i.e., $N=\left |X_t\right |$, and $l=\infty$), we discover only 40 unique addresses from  CryptoLocker (Padua), CryptoWall (Padua), CryptoTorLocker2015 (Montreal), CryptoTorLocker2015 (Padua) families. 

\begin{figure}
       \centering
       \includegraphics[width=0.9\linewidth]{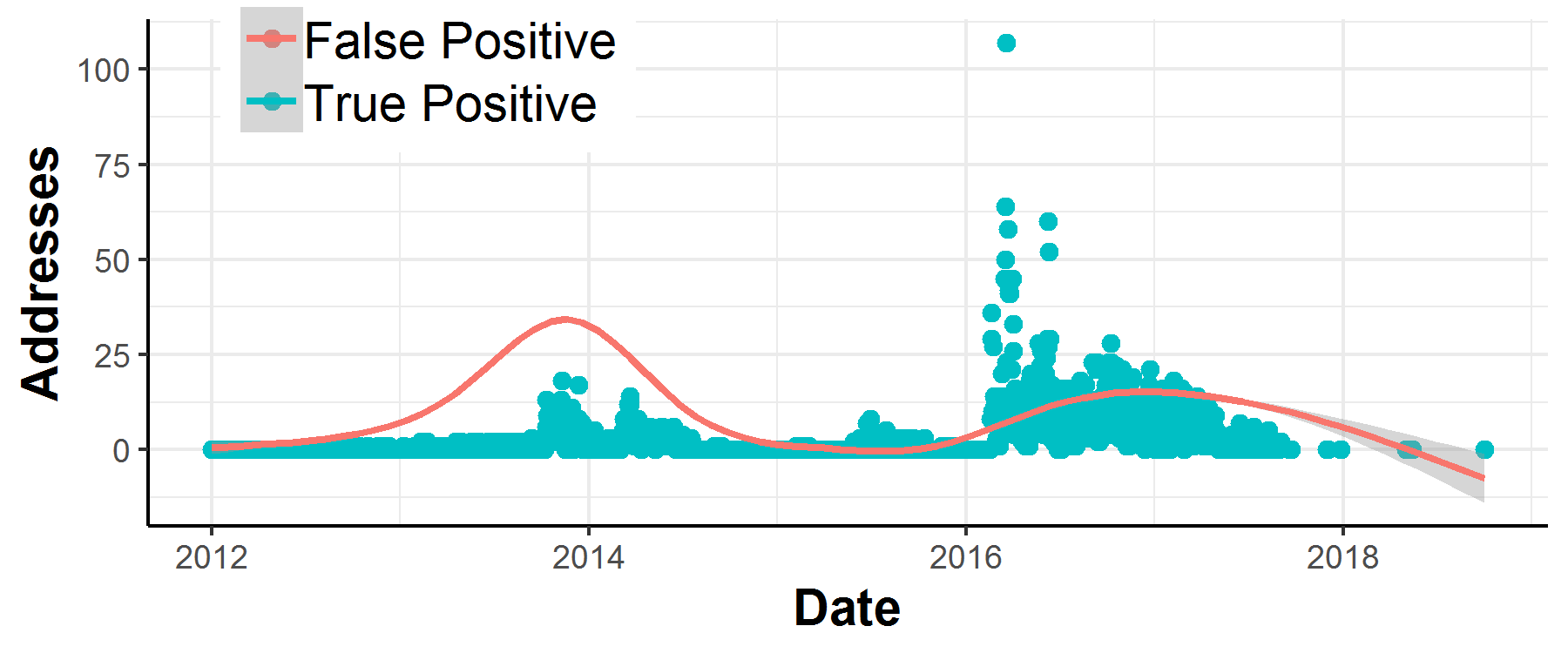}
       \caption{Ransomware detection with exact feature matches.}
       \label{fig:naive}
 \end{figure}

\xhdr{Naive similarity search.} An interesting benchmark for detecting undisclosed payments from existing families is to compute the similarity of $X_{t^\prime}$ addresses to the past addresses in $X_{t}$. If existing families exhibit repeating patterns over time, the similarity search can match new addresses to known ransom addresses. In fact, Fig.~\ref{fig:naive} shows that this strategy may be effective. By using exact matches to known ransom patterns of the past, in six days similarity search reveals more than 50 addresses each day, while predicting a maximum of 73 false positives on 2013 day 336. These results offer supporting evidence of utility of our features. However, this naive approach creates 21,371 FP addresses overall, which makes it unfeasible for operational use by security analysts.  

\begin{table}[htbp]
  \centering
  \caption{Existing family address detection results. }
    \setlength\tabcolsep{1.3pt}
    \begin{tabular}{llrrrrrrrrrrr}
    RS & Method  & {$l$} & {$N$} & {TP} & {FP} & {FN} & {TN} & {\#w} & {Prec} & {Rec} & {F1} & {PLR} \\ \hline
    
    Locky & $\TDA{.9}{.8}{.5}$& 240   & 300   & 451   & 2350  & 50    & 8221  & 11    & 0.161 & 0.900 & 0.273 & 0.192 \\
     & $\mathbb{COSINE}$ & 90    & 300   & 2395  & 41681 & 3990  & 146369 & 194   & 0.054 & 0.375 & 0.095 & 0.057 \\ \hline
    Crypto &$\TDA{.9}{.8}{.65}$   & 240   & 600   & 217   & 3087  & 155   & 11200 & 15    & 0.066 & 0.583 & 0.118 & 0.070 \\
    Wall &$\mathbb{DBSCAN}_{.2}$ & 240   & 600   & 728   & 18960 & 794   & 16913 & 59    & 0.037 & 0.478 & 0.069 & 0.038 \\ \hline 
    Crypto &$\TDA{.9}{.65}{.65}$   & 240   & 300   & 439   & 9686  & 212   & 22129 & 34    & 0.043 & 0.674 & 0.081 & 0.045 \\
     Locker& $\mathbb{DBSCAN}_{.15}$ & 60    & 300   & 935   & 42771 & 295   & 11316 & 67    & 0.021 & 0.760 & 0.042 & 0.022 \\ \hline
    Cerber & $\TDA{.9}{.5}{.35}$ & 120   & 300   & 187   & 5174  & 459   & 23027 & 29    & 0.035 & 0.289 & 0.062 & 0.036 \\
     & $\mathbb{XGBOOST}$ & 240   & 300   & 1606  & 47307 & 7279  & 374169 & 436   & 0.033 & 0.181 & 0.056 & 0.034 \\ \hline
    Crypt & $\TDA{.9}{.35}{.35}$ & 90    & 300   & 77    & 2460  & 271   & 11057 & 14    & 0.030 & 0.221 & 0.053 & 0.031 \\
    XXX & $\mathbb{COSINE}$ & 30    & 600   & 589   & 20872 & 610   & 42952 & 65    & 0.027 & 0.491 & 0.052 & 0.028 
     \end{tabular}%
  \label{tab:1aresults}%
\end{table}%
Table~\ref{tab:1aresults} shows the main results of our models. For each ransomware family, TDA has a hyper-parameter set that produces the best model. However, these parameter values are not the same across families. Sample size ($N$) and training length ($l$) parameters are different as well. For each family, we also provide the best non-TDA model for comparison. The Locky ransomware has the best results with a precision of 0.161 and recall of 0.9. In general, models yield better recall than precision. In Table~\ref{tab:1aresults}, $\# w$ is the number of windows where a model makes at least one label prediction. By using the $q,\epsilon_1$ and $\epsilon_2$ hyper-parameters, TDA models avoid predicting labels when level of confidence in the derived classification is low.

Similar to TDA, DBSCAN can ignore data points in clustering; two of the best non-TDA results are delivered by DBSCAN models. In the best TDA models for each ransomware family, \textbf{we predict 16.59 false positives for each true positive}. In turn, this number is 27.44 for the best non-TDA models.

 \begin{figure}
       \centering
       \includegraphics[width=0.9\linewidth]{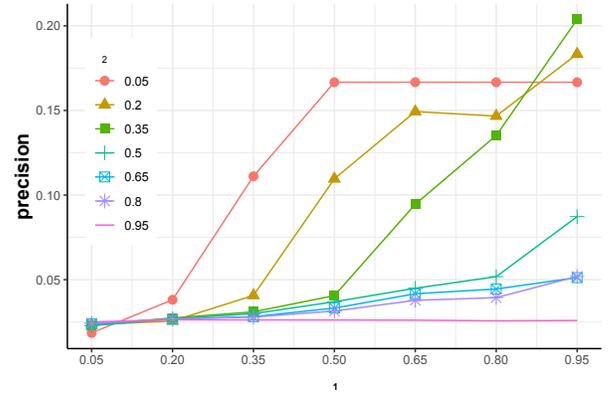}
       \caption{$\epsilon$ parameters in all TDA models (the filter threshold $q=0.9$).}
       \label{fig:1ahyper}
\end{figure}

As TDA models appear to be sensitive to the selected hyper-parameters, we turn our attention to these values. For increasing threshold values, recall decreases from 0.53 ($q=0.0$) to 0.37 ($q=0.9$), but precision and PLR values do not change.  In Figure~\ref{fig:1ahyper}, we present model performance for $\epsilon_1$ (i.e., threshold on how many past ransomware addresses should be contained in the cluster) and $\epsilon_2$ (i.e., threshold on at most how many addresses the cluster can contain). As Figure~\ref{fig:1ahyper} indicates, TDA models tend to be more sensitive to $\epsilon_2$ values. By limiting $\epsilon_2$ to small values, we reach higher precision on average. However, this leads to fewer predictions, as $\epsilon_2$ defines at most how many data points can be grouped together (see sec.~\ref{sec:tda}).

 \begin{figure*}[ht!]
 \centering
\begin{subfigure}{.33\textwidth}
 \includegraphics[width=1.0\linewidth]{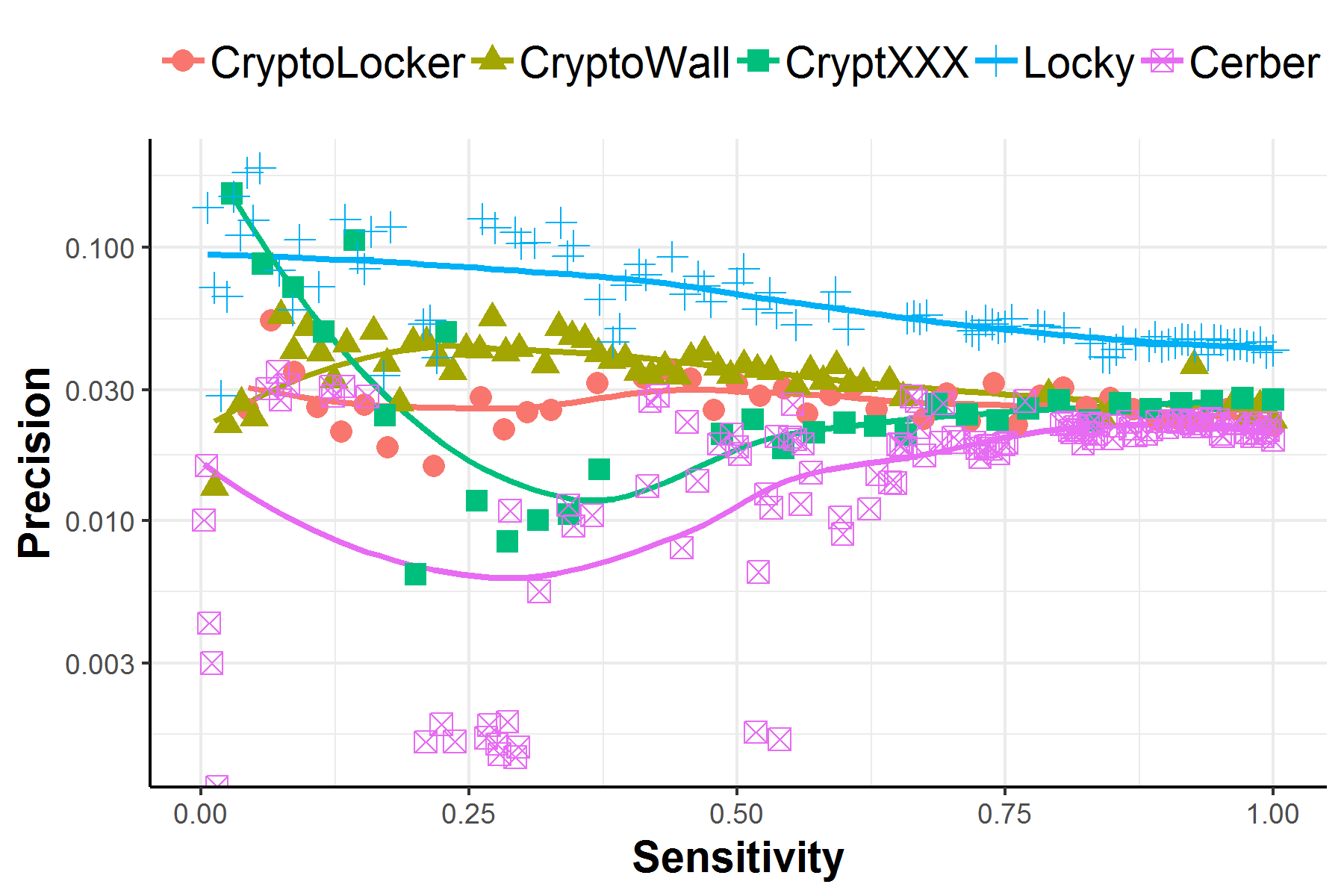}
\end{subfigure}%
~
\begin{subfigure}{.33\textwidth}
   \includegraphics[width=1.0\linewidth]{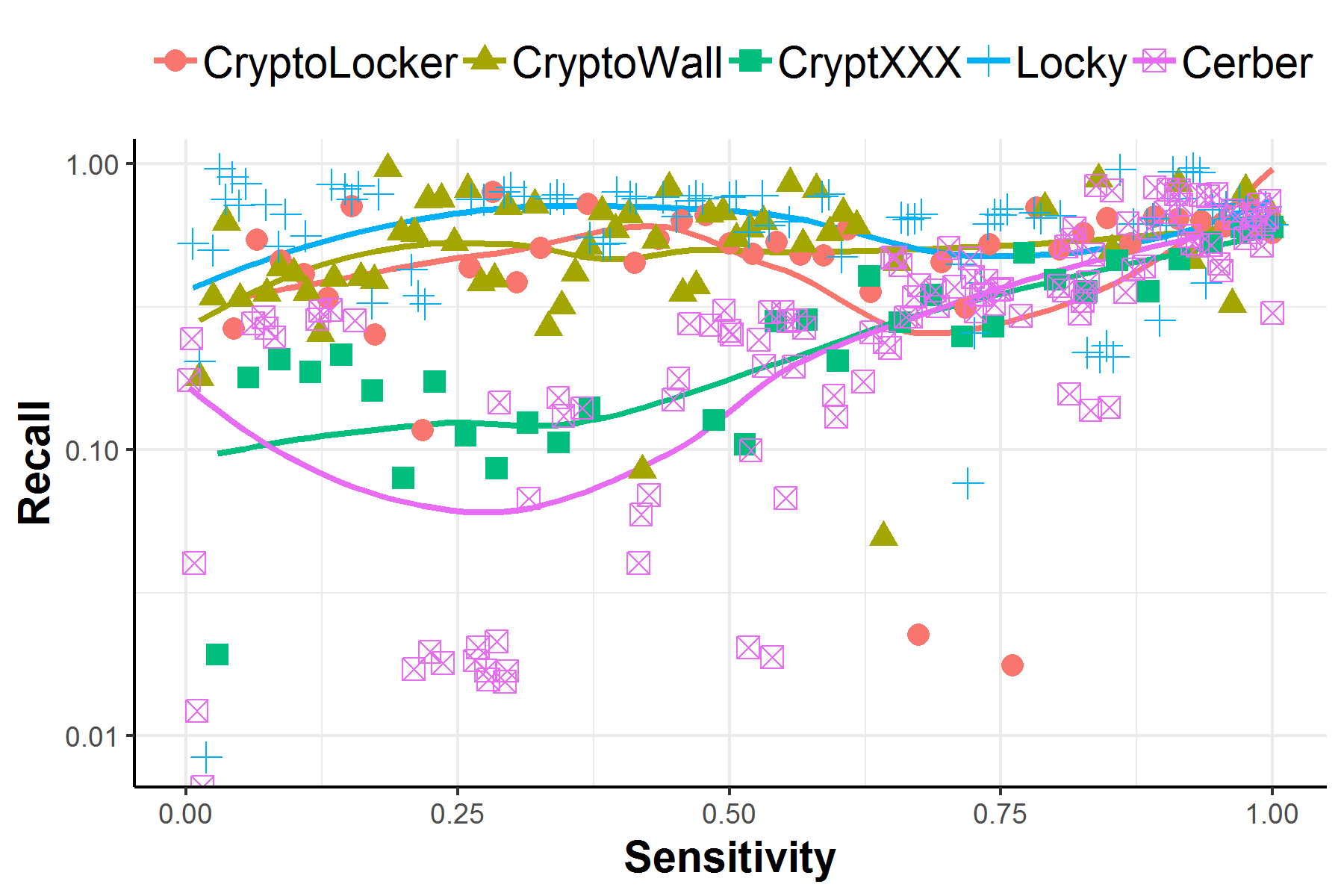}
\end{subfigure}%
~
\begin{subfigure}{.33\textwidth}
   \includegraphics[width=1.0\linewidth]{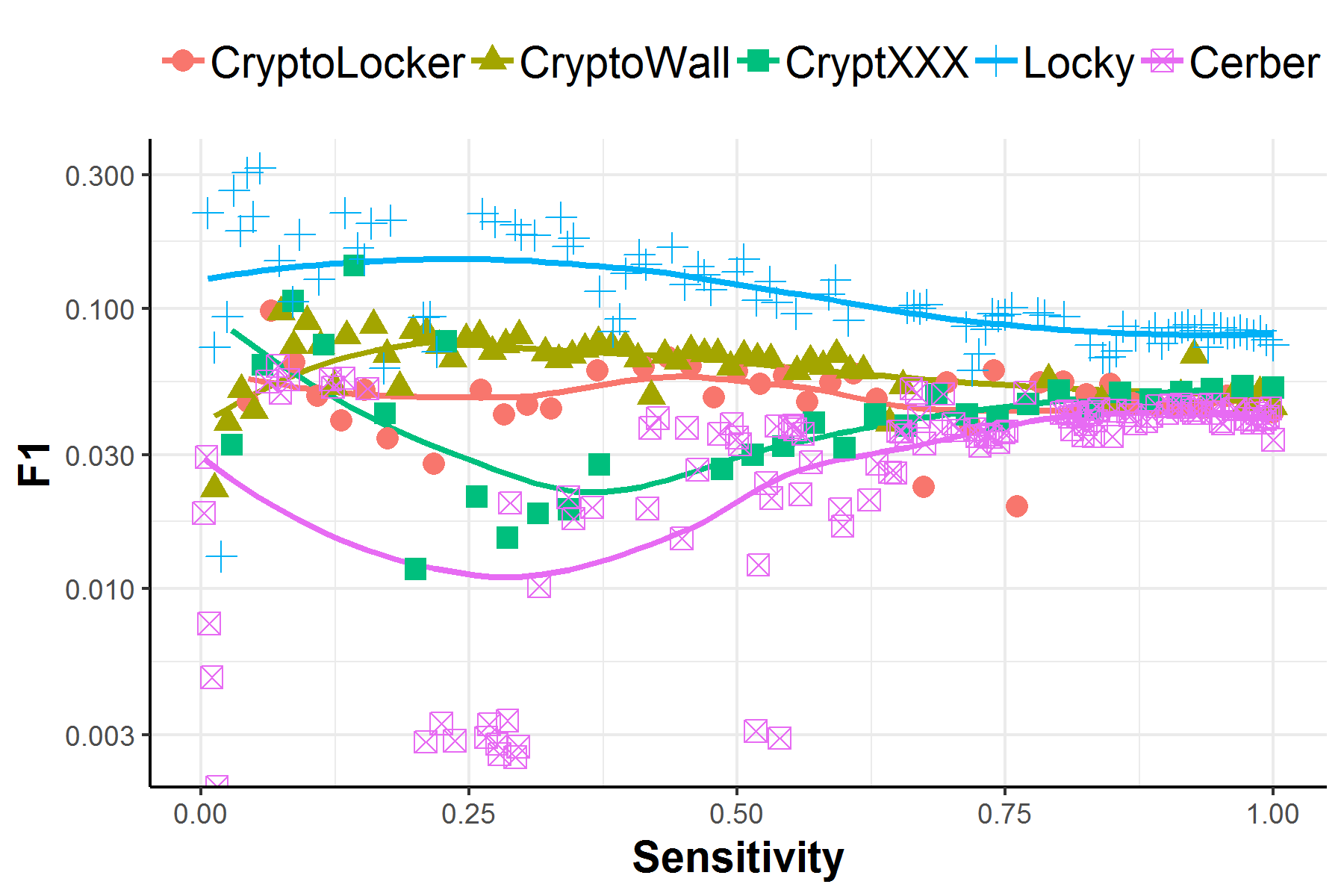}
\end{subfigure} 
\caption{Performance of TDA models with increasing sensitivity. Sensitive models can make predictions for every window. }
\label{fig:1a1sensitivity}
\end{figure*}

Our results indicate that making predictions in every time window may not be desirable, as it leads to lower F1 values. Both TDA and DBSCAN can ignore some data points, thereby making predictions only when \textit{enough} data exists. As we can control the sensitivity of TDA models through hyper-parameter selection, a natural question is to ask whether more restrictive settings should be preferred in TDA.  Figure~\ref{fig:1a1sensitivity} depicts TDA performance when settings allow for lower (i.e., fewer windows) or higher (i.e., more windows) sensitivity. Precision, recall and F1 values are averaged for TDA models. We find that less sensitive models yield higher precision, but lower recall values. Cerber results show the greatest change in terms of sensitivity, whereas Locky has the best and most stable results. For some families, such as CryptXXX, the least sensitive models deliver the highest precision values. These findings imply that some TDA models are able to make predictions for very few windows only, but the results can be very accurate.  

\subsection{New Family Prediction}

We outline our experimental settings for the detection problem as follows. 
Given features $X_t$ and known labels $Y_t$ of past addresses at time $t$ and features of addresses $X_{t^\prime}$ at time $t^\prime >t$, we train for discovering addresses belonging to new ransomware families as follows:
\begin{enumerate}[leftmargin=0.1cm,itemindent=.5cm,labelwidth=\itemindent,labelsep=0cm,align=left] 
    \item For $t< t^\prime$, use a training length $l$, and create a dataset $X_{\left[t-l,t\right]} \subseteq X_t$ which holds features of addresses observed between windows $t-l$ and $t$. 
    \item Create an $f_0$ sample of size N from $X^0_{\left[t-l,t\right]} \subseteq X_{\left[t-l,t\right]}$ without replacement where $\forall x_u \in X^0_{\left[t-l,t\right]} , y_u=f_0$ and $N=\left |X^0_{\left[t-l,t\right]}\right|$. 
    \item Create a \textit{ransomware} sample of size N from $X^{rs}_{\left[t-l,t\right]} \subseteq X_{\left[t-l,t\right]}$ without replacement where $\forall x_u \in X^{rs}_{\left[t-l,t\right]} , \textcolor{red}{y_u\neq f_0}$ and $N \leq \left |X^{rs}_{\left[t-l,t\right]}\right|$. 
    \item \label{l2:1} Relabel all addresses in  $Y^{rs}_{\left[t-l,t\right]}$ with the label $f_r$.
    \item By using the ground truth data at $t^\prime$, take a sample of M=1000 white addresses without replacement to be used in the testing phase: $X^0_{t^\prime}$.
   \item By using the ground truth data, choose a family $rs^\prime$ whose emergence at $t^\prime$ will be discovered.
    \item By using the ground truth data at $t^\prime$, find all ransomware addresses for $t^\prime$ to be used in the testing phase:  $X^{rs^\prime}_{t^\prime}$.
    \item \label{l2:2} Use features $\{ X^0_{\left[t-l,t\right]} \cup X^{r}_{\left[t-l,t\right]} \}$ and labels $\{ Y^0_{\left[t-l,t\right]} \cup Y^{r}_{\left[t-l,t\right]} \}$ as the training data, and classify $\{ X^0_{t^\prime} \cup X^r_{t^\prime} \}$.
\end{enumerate}
 
\noindent We emphasize two aspects in predicting a new family: i) [Step~\ref{l2:1}]: in training, addresses of all existing families are relabeled with $f_r$, thereby creating a unified ransomware class, ii) [Step~\ref{l2:2}]: when an address is predicted as ransomware, we cannot immediately claim that it constitutes a new family, or belongs to an existing family.

As our current goal is to predict new ransomware families without any prior information about these families, \textit{the training task has to learn a single model that will be used to identify all future ransomware families}. Using our models, we forecast emergence of 25 ransomware families. Emergence of the first ransomware, CryptoLocker, cannot be predicted, because we have no prior ransomware data to train a model for CryptoLocker.  
 
The best model, $\TDA{0.7}{0.05}{0.35}$  uses $N=1000$ past samples, $l=120$ training length, and predicts $TP=26, FN=8, TN=5032, FP=21075$. The model predicts 25 emerging ransomware families, but also \textbf{results in 810.57 false positives  for each true positive}.  

We hypothesize that this unsatisfactory performance is due to data scarcity; among the 25 families, only three families have more than one address in their first window on the Bitcoin blockchain. These families are DMALockerv3 (2016 day 233), Flyper (2016 day 335) and eRanger (2016 day 68).

\begin{table}[htbp]
  \centering
  \caption{Day/Year pairs in the discovery experiment.}
    \begin{tabular}{lrrr}
Ransomware&First window&Used window&\#Unique address\\ 
\hline
Cerber&62/2016&89/2016&16\\
CryptXXX&132/2016&133/2016&38\\
DMALocker&7/2015&34/2016&14\\
CryptoWall&59/2014&64/2014&22\\
Locky&42/2016&47/2016&59
    \end{tabular}%
  \label{tab:dates}
\end{table}%

We now repeat the forecasting experiment by considering the \textit{earliest} window when a ransomware yields 10 or more addresses. Such filtering results in a dataset of five ransomware families.  We exclude the previously observed addresses of these families from our training set. The first and the 
identified earliest windows for each family are presented in Table~\ref{tab:dates}. Time difference in windows is as small as one for some families.

\begin{table}[htbp]
  \centering
  \caption{Model performance for new family prediction.}
    \begin{tabular}{lrrr}
Method&Training size $N$ &Training length $l$&{PLR}\\
\midrule
$\TDA{0.9}{0.2}{0.2}$&300&60&0.21085470\\
$\mathbb{COSINE}$&300&60&0.04232920\\
$\mathbb{DBSCAN}_{0.2}$&300&60&0.02802958\\
$\mathbb{RANDOM FOREST}$&300&60&0.00000000\\
$\mathbb{XGBOOST}$&300&60&0.00000000

    \end{tabular}%
  \label{tab:models}
\end{table}%
   
Despite improving data scarcity, Table~\ref{tab:models} shows that tree based methods (i.e., Random Forest and XGBoost) fail to predict any ransomware family. Table~\ref{tab:2BBest} shows predicted values for each family. We find that three TDA models tend to deliver the best F1 results for all five families. In addition to TDA models, we show one competing result from other models for each family. Overall, in three families TDA has the highest PLR value. For DMALocker, the best models are DBSCAN-based ones. Unlike existing family detection, in new family prediction we value finding an address with the fewest false positives. In this aspect, we reach the best result for CryptXXX, where a TDA model predicts one true positive and one false positive. With the best models provided in Table~\ref{tab:2BBest}, on average \textbf{we predict 27.53 false positives for each true positive} in forecasting of new ransomware families.   

However, we emphasize that for some families, such as CryptXXX, our models predicts only two ransomware addresses, one of which is a true positive. This result offers evidence that our prediction models can be highly effective for certain families of ransomware.

\begin{table}[htbp]
  \centering
  \caption{Results for new ransomware family prediction ($l=60$, $N=300$).}
  \setlength\tabcolsep{2pt}
    \begin{tabular}{llrrrrrrrr}
RS&Method&Prec&Rec&TN&FP&TP&FN&PLR\\
\midrule
CryptXXX&$\TDA{0.9}{0.2}{0.2}$&0.500&0.026&917&1&1&37& 1.0\\
&$\mathbb{COSINE}$&0.046&0.342&654&264&13&25&0.049\\
\midrule
Locky&$\mathbb{COSINE}$&0.098&0.138&795&37&4&25&0.108\\
&$\TDA{0.9}{0.05}{0.95}$&0.047&0.586&489&343&17&12&0.049\\
\midrule
CryptoWall&$\TDA{0.9}{0.05}{0.95}$&0.0625&0.500&810&165&11&11&0.067\\
&$\TDA{0.9}{0.35}{0.8}$&0.061&0.500&805&170&11&11&0.0647\\
\midrule
Cerber&$\TDA{0.9}{0.05}{0.95}$&0.029&0.214&849&100&3&11&0.030\\
&$\TDA{0.9}{0.35}{0.8}$&0.023&0.642&570&379&9&5&0.023\\
\midrule
DMALocker&$\mathbb{DBSCAN}_{0.2}$&0.019&0.875&120&367&7&1&0.019\\
&$\mathbb{DBSCAN}_{0.15}$&0.015&0.875&4&459&7&1&0.015
    \end{tabular}%
  \label{tab:2BBest}
\end{table}%

\begin{table}[htbp]
  \centering
  \caption{Mean values of metrics for the filter threshold in TDA models across five families.}
    \begin{tabular}{lrrrr}
$q$&PLR&Precision&Recall&TP\\
\midrule
0.0&0.028&0.0264&0.664&15.25\\
0.3&0.028&0.0266&0.635&14.61\\
0.5&0.029&0.0278&0.589&13.75\\
0.7&0.031&0.0298&0.515&12.15\\
0.9&0.035&0.0319&0.352&8.35
    \end{tabular}%
  \label{tab:qthreshold}
\end{table}%

\begin{figure}[ht!]
 \centering
\begin{subfigure}{.48\textwidth}
 \includegraphics[width=1.0\linewidth]{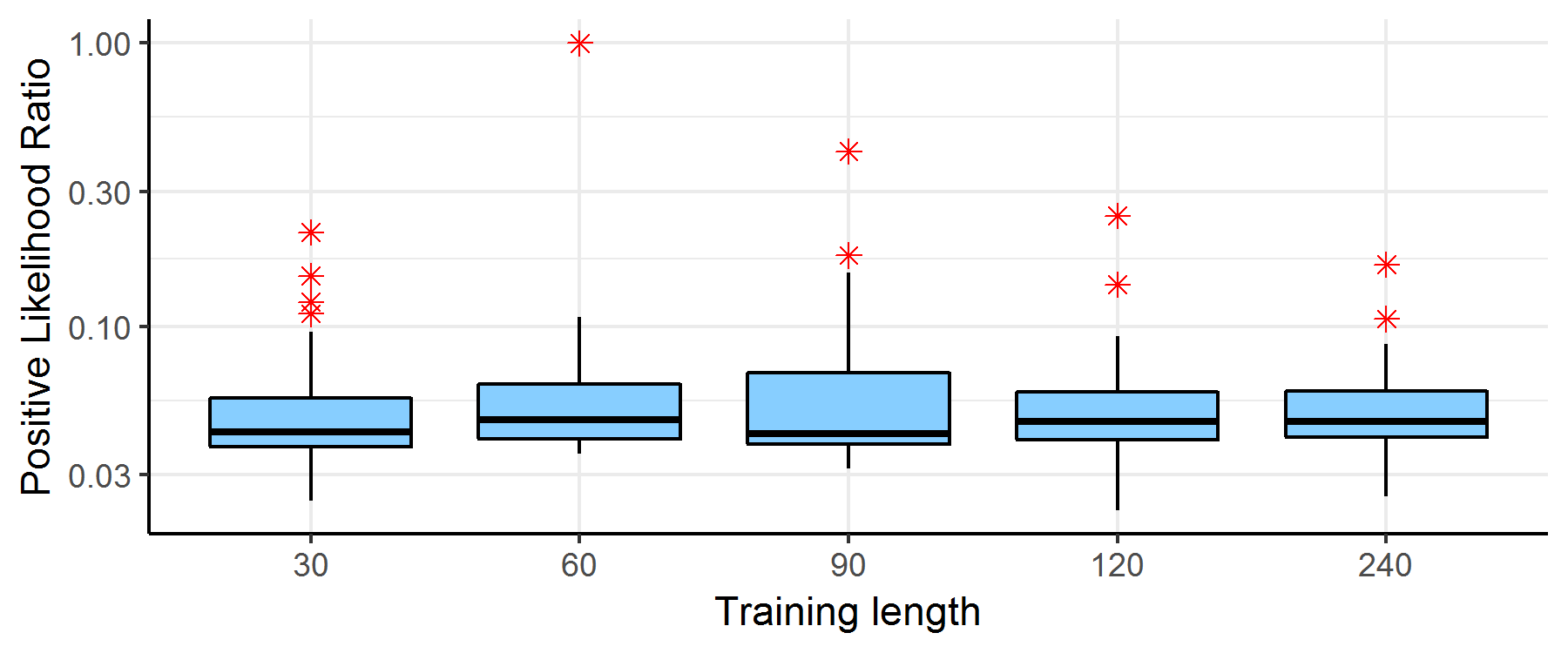}
\end{subfigure}%
\\
\begin{subfigure}{.48\textwidth}
   \includegraphics[width=1.0\linewidth]{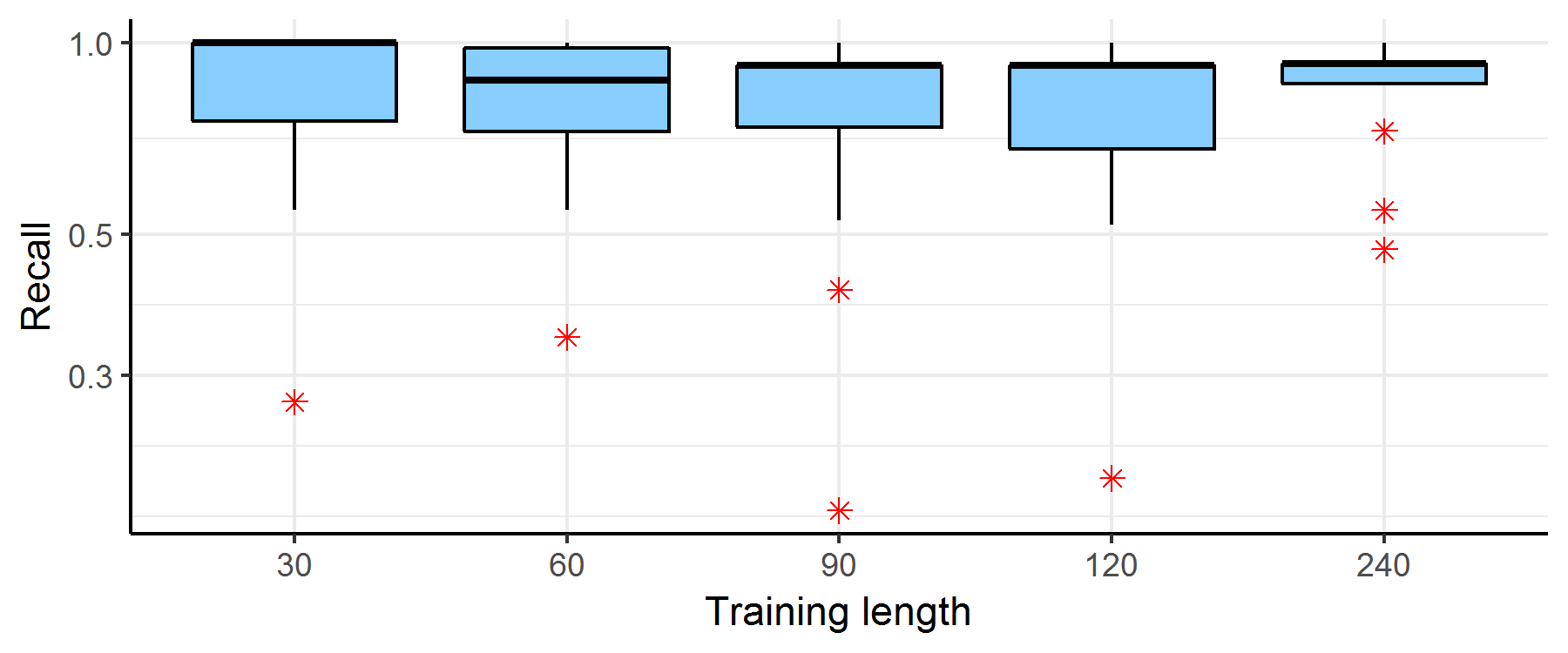}
\end{subfigure}%
\caption{Model parameters and performance values  across best models of the five ransomware families.}
\label{fig:lookpast}
\end{figure}

\begin{figure}[ht!]
 \centering
\begin{subfigure}{.48\textwidth}
 \includegraphics[width=1.0\linewidth]{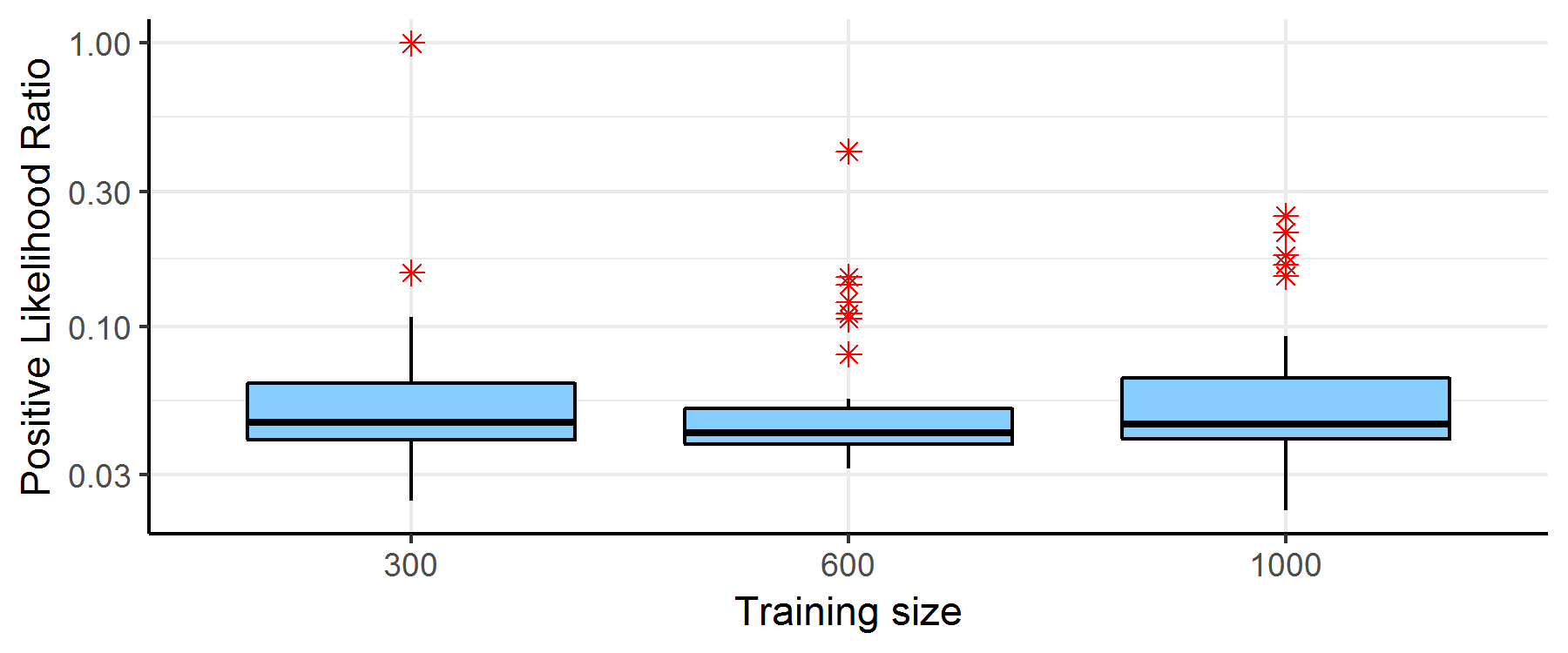}
\end{subfigure}%
\\
\begin{subfigure}{.48\textwidth}
   \includegraphics[width=1.0\linewidth]{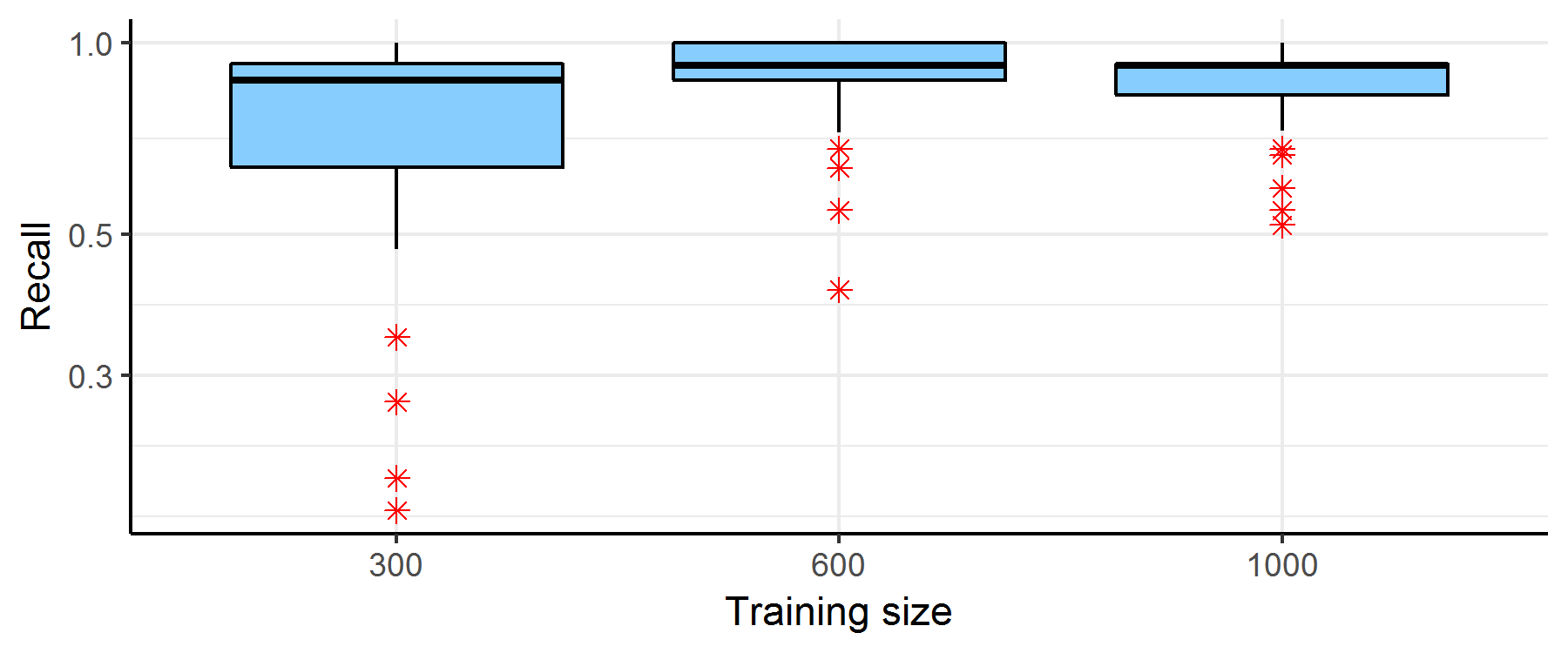}
\end{subfigure}%
\caption{The impact of training size across best models of the five ransomware families.}
\label{fig:trainingsize}
\end{figure}

\xhdr{Model parameters in prediction.} Table~\ref{tab:qthreshold} shows average model performance across five ransomware families for different $q$ values in TDA models.  As the threshold increases, we observe slightly higher precision and lower recall values.  Figures~\ref{fig:lookpast} and~\ref{fig:trainingsize} depict PLR and recall values for each family. Overall, families exhibit better results for smaller training lengths (i.e., $l$) and larger sample size (i.e., $N$) values. This implies that ransomware addresses change features in time, and using a long history does not always result in better model performance.

\section{Address Elimination}
\label{sec:elimination}

As a final step in our ransomware prediction, we employ three filtering approaches on the results produced by the best models in the existing family detection task.

\xhdr{Setting.} To this end, we use the best TDA models outlined in Table~\ref{tab:1aresults} with $N=300$, $l=60$, and classify addresses in the last 30 days of each ransomware family. This task results in 72 days, where we can forecast and test the predictive performance.   CryptoLocker, CryptXXX, DMALocker, Cerber and Locky have 2, 20, 1,19 and 20 days' predictions, respectively.  In these forecasting tasks, 3,211 addresses have been labeled as \textit{suspicious} (potential ransomware), whereas only 92 of them are true ransomware addresses.  

\xhdr{Shape Elimination.} We hypothesize that ransomware transactions can be divided into two types: front and mixing. Front transactions are initial payments made by ransomed entities. Mixing transactions are created by hackers to transfer and sell received coins to other addresses. The ransomed entities buy bitcoins from exchanges or bitcoin users to pay hackers. These transactions are N-to-1 (i.e., in a transaction that has N input addresses and one output address, all coins are gathered in one address) or N-to-2 (i.e., a remaining amount is further directed to a change address) transactions. In this setting, depending on how many bitcoins are bought, N can be as high as 15000. 

If we focus on detecting front payments only, we hypothesize that number of ransomware labeled addresses can be further reduced.  On the complete Bitcoin blockchain, 91.06\% of all transactions are N-to-1 or N-to-2 transactions. Of the 24198 known ransomware addresses in our dataset, 20826 (86.06\%) have at least one appearance that receives coins in an N-to-1 or N-to-2 transactions. Of the 3,211 addresses, we find that 861 (26.8\%) can be filtered out. However, 16 of the 92 true ransomware addresses are also filtered out. \textbf{Shape elimination results in 2,350 suspicious addresses and 76 ransomware addresses}. 

\begin{figure}
       \centering
       \includegraphics[width=0.9\linewidth]{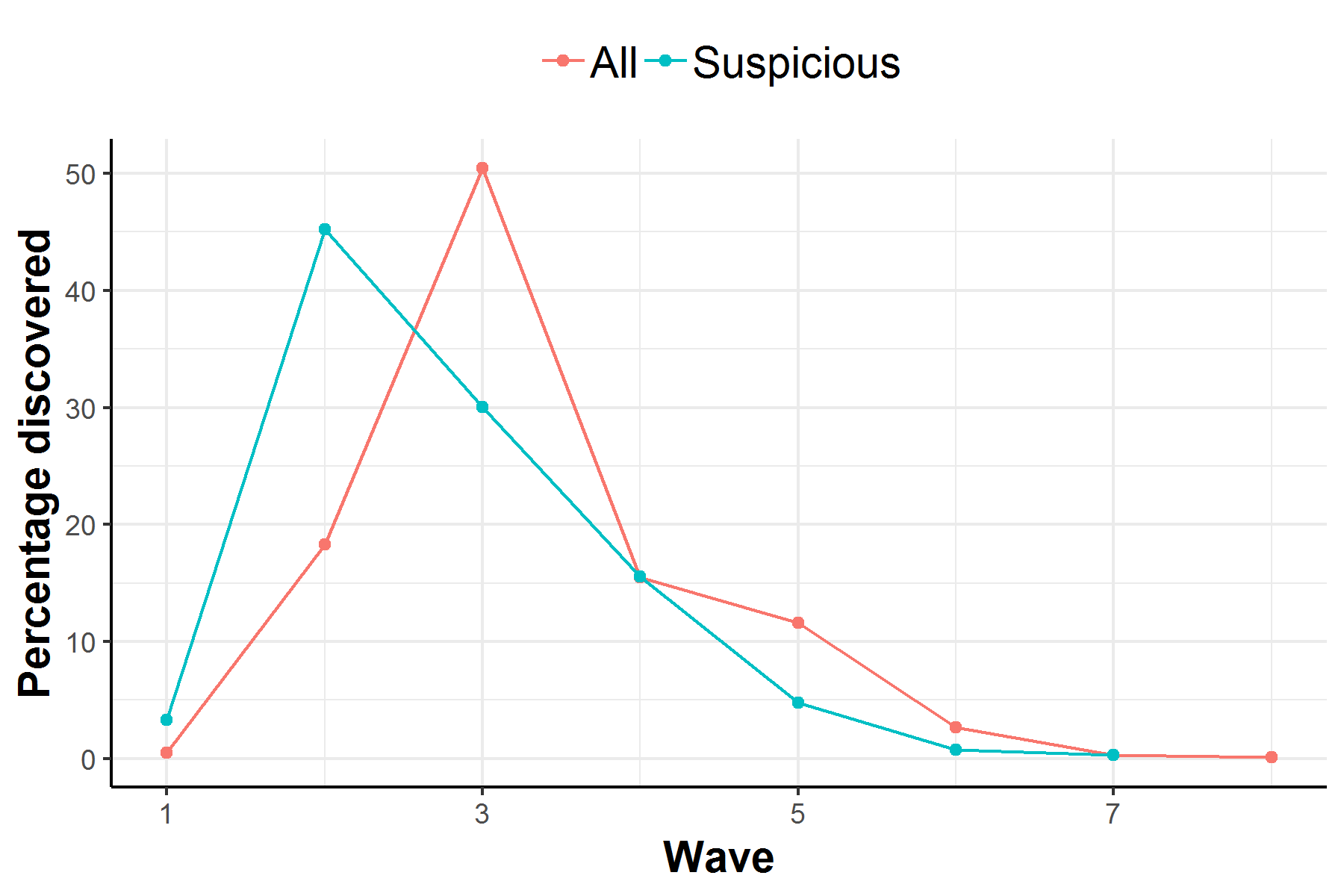}
       \caption{Distance of ransomware labeled (suspicious) addresses and all addresses to known ransomware addresses. Suspicious addresses are closer to known ransomware.   Difference of the two distributions is statistically significant as indicated by Kolmogorov-Smirnov test~\cite{lilliefors1967kolmogorov} with $D = 0.29564$, and $p$-value $< 2.2\times 10^{-16}$.}
       \label{fig:4ADist}
   \end{figure}

\xhdr{Graph Elimination.} Once ransom is received, hackers can sell each ransom coin to a different buyer (address) individually. This approach makes ransomware address detection particularly challenging. However, in most cases transaction costs discourage this privacy enhancing  approach, and coins are merged and sold together in a single transaction.  For example, Huang et al. \cite{huang2018tracking} found that coins of Cerber family were continuously merged, and the resulting large tree of addresses was  easily identified by the researchers. 

We hypothesize that the small distance between an address to known past ransomware addresses can help us to be more confident in its label designation. We measure the distance in a breadth first search as the number of edges on the shortest path to a known ransomware address. Figure~\ref{fig:4ADist} depicts the distribution of 666 addresses 
which are reachable from the known past virus addresses on the graph. Here, 8 of the 92 true ransomware addresses are contained in these 666 addresses. Furthermore, the 8 unique addresses were at most 4 hops away from known ransomware addresses. 

Figure~\ref{fig:4ADist} indicates that among all addresses that are reachable from the known ransomware addresses, the suspicious addresses tend to be located noticeably closer. Mean and median distance values of suspicious nodes to ransomware nodes are $\bar{x}=2.44$, $\tilde{x}=3$, respectively, with a maximum distance of 7 hops. If we use a maximum distance of 4 or less hops, only 617 addresses remain.  \textbf{Graph elimination hence results in 617 suspicious addresses and 8 true ransomware addresses.}

\xhdr{Corroborating evidence:} In all ransomware families, we record 324 addresses that are found to be suspicious in multiple time windows. In maximum, 3 addresses appear in 6 time windows each. All 324 addresses survive shape elimination, whereas only 158 addresses survive graph elimination. In total, we find \textbf{350 unique suspicious addresses and 8 true ransomware addresses after graph and shape eliminations}.

\xhdr{Multisig elimination.} Bitcoin contains multiple types of addresses. In the simplest case, an \textit{ordinary} address is controlled by a single private key. These addresses start with the character '1' such as {\tiny 1LXrSb67EaH1LGc6d6kWHq8rgv4ZBQAcpU}. A second address type, \textit{multisig address}, requires more than one key to authorize a Bitcoin transaction. Multisig addresses start with the character '3' such as {\tiny 3NQoq5MVPfEMw12gB4a2c1G61mRZyMymsB}. These addresses are usually used to divide up responsibility for possession of bitcoins. 
For the first usage, as ransomware payments involve a payer and a payee who may not trust each other, we advocate that multisig addresses cannot  be used in ransomware payments. 

A second usage is to prevent coin theft: users can create a multisig transaction where they own all the addresses, but private keys are stored on different machines. In such schemes, coins can only be stolen if all machines are compromised, which makes the theft very difficult. Hackers could potentially use multisig addresses to avoid theft of ransomed coins. Though, we believe that such a scenario is currently very rare, since only three addresses with multisig are found among the more than 24K known ransomware related addresses. Of course, this heuristic may not be effective in the future, if ransomware operators start using more multisig addresses.    

Out of the 358 suspicious address, 339 are ordinary addresses; these addresses start with '1'. Hence, our multisig elimination round ends with 8 true ransomware, and 331 suspicious addresses. 

Our elimination results show that many addresses are found suspicious in multiple time windows. Furthermore, such addresses survive all graph, chainlet and multisig eliminations. \xhdr{Our results offer strong evidence that these suspicious addresses are indeed ransomware related.} 
 
 \section{Conclusions}
In this paper, we propose a new data analytics based framework to detect and predict ransomware transactions that use Bitcoin. 
Using a topological data analysis based approach and novel blockchain graph related features, we empirically show that we can significantly improve the ransomware related Bitcoin address detection accuracy compared to existing heuristic based approaches. Furthermore, we show that by using certain address elimination heuristics, we can reduce the number of suspicious addresses that need to be analyzed by the cyber security analysts. 

As a future work, we plan to combine our proposed approach with other methods (e.g., running the ransomware to get additional seed addresses) and threat intelligence information (e.g., reports about the emergence of new ransomware) to increase our prediction accuracy.

\begin{table*}[htbp]
  \centering   
  \caption{Symbols and their explanations.}
  \small
    \begin{tabular}{rcr}
    
    Symbol&Explanation&Section\\
    \midrule
$\mathcal{G}, V, E, B$ & Graph, its nodes, edges and node types& \ref{sec:bitcoingraph}\\
$TX$& set of all transactions&\\
$\phi(u)$&node type of $u$&\\
$\Gamma_a^i$ and $\Gamma_a^o$ & in and out neighbors of $a$& \\
$a_u$, $\vec{x}_u$, $y_u$&address, address feature vector, address label&\ref{sec:methodology} \\
D&number of features& \\
$f_o$,$f_i$& non-rs address, address of ransomware family $i$& \\
$X_t$, $X_{t^\prime}, Y_t$& training features, test features, training labels&\\
$tx_n, \mathcal{A}_o(n), A^o_u(n)$&transaction n, sum of its outputs, output of $tx_n$ to address u&\\
$a_u^t$&appearance time t of address u&\\
$I_u$, $L_u$,$W_u$, $C_u$, $O_u$&income, length, weight, count, loop of address u&\\
$\mathcal{TX}$&starter transactions&\\
$U$, $\xi$& number of observed addresses, filter function&\ref{sec:TDAMapper}\\
$q,\epsilon_1,\epsilon_2$&quantile filter, inclusion threshold, size threshold&\\
$I$, $S$, $\{u_{jk}\}$&range,set of intervals, TDA clusters&\\
$\mathcal{CT}$, $C_c$& a cluster tree graph, a cluster in the graph&\\
$A_c$, $V$& addresses in cluster $C_c$, known past RS addresses in cluster $C_c$&\\
$P$,$S$&scores of addresses, set of suspicious addresses&\\
$l$, $N$, $M$&training lengths, sample size in training, sample size in testing& \ref{sec:experiments}\\
$X_t^0$, $X_t^{rs}$& at time t training features of non-RS addresses, rs addresses&\\
$w$ &a specific time window&\\

    \end{tabular}%
  \label{tab:symbol}%
\end{table*}%

\begin{table*}[htbp]
  \centering
  \caption{Mean feature values of ransomware families. }
   \begin{tabular}{lrrrrrrrr}
    Ramsomware &  {Weight} & {Length} & {Looped} &  {Income} &  {Neighbor} &  {Count} &  {\#address} &  {\# Unique address} \\
    \midrule 
    APT & 0.71 & 67.63 & \cellcolor{shade}734.09 & 371987303 & 2.54 & 2047  & 11    & 2 \\
    Cerber & 0.32 & 39.90 & 54.40 & 103224640 & 2.01 & 737.08 & 9223  & 9177 \\
    ComradeCircle & 0.05 & \cellcolor{shade}144.00 & 0     & 203320001 & 2 & 1241  & 1     & 1 \\
    CryptConsole & 0.59 & 43.42 & 0     & 45463341 & 2 & 831.71 & 7     & 5 \\
    CryptoLocker & 0.88 & 30.67 & 100.98 & 1840825184 & 2.88 & 308.32 & 9315  & 1509 \\
    CryptoTorLocker2015 & 1.19 & 20.58 & 166.63 & 680784614 & 10.21 & 220.36 & 55    & 32 \\
    CryptoWall & 0.79 & 47.92 & 121.01 & 701610513 & 2 & 425.87 & 12390 & 1894 \\
    CryptXXX & 0.37 & 47.44 & 61.02 & 135534326 & 2.01 & 791.84 & 2419  & 1354 \\
    DMALocker & 0.97 & 38.82 & 430.48 & 889427618 & 1.85 & 897.16 & 251   & 21 \\
    DMALockerv3 & 0.54 & 37.71 & 94.55 & 610589614 & 1.16 & 962.88 & 354   & 143 \\
    EDA2 & 0.17 & 74.33 & 951.5 & 37852094 & 1.33 & \cellcolor{shade}4355.16 & 6     & 4 \\
    Flyper & 0.51 & 19.11 & 0.00     & 49777364 & 1.44 & 324.22 & 9     & 8 \\
    Globe & 0.48 & 56.06 & 393.65 & 79555251 & 2.09 & 1623.90 & 32    & 7 \\
    GlobeImposter & 0.51 & 26.47 & 54.87 & 1076278167 & 1.58 & 357.2 & 55    & 1 \\
    Globev3 & 0.45 & 71.82 & 151.44 & 119933550 & 2.12 & 1377.29 & 34    & 5 \\
    KeRanger & 0.36 & 50.20 & 280.7 & 99990000 & 1 & 1021.1 & 10    & 10 \\
    Locky & 0.37 & 46.90 & 88.65 & 244419979 & 1.25 & 1027.25 & 6625  & 6584 \\
    NoobCrypt & 0.85 & 23.19 & 80.13 & 228913211 & 1.29 & 338.23 & 483   & 28 \\
    Razy & \cellcolor{shade}43.21 & 32.62 & 199.77 & \cellcolor{shade}228205910311 & \cellcolor{shade}10.23 & 744.85 & 13    & 1 \\
    Sam & 0.06 & 6.00 & 1     & 2900000000 & 3 & 1     & 1     & 1 \\
    SamSam & 0.73 & 44.67 & 123.15 & 1078297206 & 1.55 & 1143.63 & 62    & 44 \\
    VenusLocker & 0.09 & 29.14 & 0     & 97142857 & 1.58 & 1     & 7     & 1 \\
    WannaCry & 0.58 & 100.57 & 801.25 & 67141405 & 1.71 & 4422.36 & 28    & 5 \\
    XLocker & 0.41 &  \cellcolor{shade}144.00 & 0     & 100000000 & 1 & 4511  & 1     & 1 \\
    XLockerv5.0 & 0.32 & 44.85 & 0     & 185710813 & 1.14 & 1063.57 & 7     & 3 \\
    XTPLocker & 0.34 & 108.50 & 285.75 & 263743359 & 1.38 & 2938.88 & 8     & 3 
     \end{tabular}%
  \label{tab:summary}%
\end{table*}%

\balance  
\bibliographystyle{IEEEtranS}
\bibliography{acmart}

\begin{thebibliography}{10}
\providecommand{\url}[1]{#1}
\csname url@samestyle\endcsname
\providecommand{\newblock}{\relax}
\providecommand{\bibinfo}[2]{#2}
\providecommand{\BIBentrySTDinterwordspacing}{\spaceskip=0pt\relax}
\providecommand{\BIBentryALTinterwordstretchfactor}{4}
\providecommand{\BIBentryALTinterwordspacing}{\spaceskip=\fontdimen2\font plus
\BIBentryALTinterwordstretchfactor\fontdimen3\font minus
  \fontdimen4\font\relax}
\providecommand{\BIBforeignlanguage}[2]{{%
\expandafter\ifx\csname l@#1\endcsname\relax
\typeout{** WARNING: IEEEtranS.bst: No hyphenation pattern has been}%
\typeout{** loaded for the language `#1'. Using the pattern for}%
\typeout{** the default language instead.}%
\else
\language=\csname l@#1\endcsname
\fi
#2}}
\providecommand{\BIBdecl}{\relax}
\BIBdecl

\bibitem{akcora2018forecasting}
C.~G. Akcora, A.~K. Dey, Y.~R. Gel, and M.~Kantarcioglu, ``Forecasting bitcoin
  price with graph chainlets,'' in \emph{Pacific-Asia Conference on Knowledge
  Discovery and Data Mining (PaKDD)}.\hskip 1em plus 0.5em minus 0.4em\relax
  Springer, 2018, pp. 765--776.

\bibitem{heldroid}
N.~Andronio, S.~Zanero, and F.~Maggi, ``Heldroid: Dissecting and detecting
  mobile ransomware,'' in \emph{Research in Attacks, Intrusions, and Defenses},
  H.~Bos, F.~Monrose, and G.~Blanc, Eds.\hskip 1em plus 0.5em minus 0.4em\relax
  Cham: Springer International Publishing, 2015, pp. 382--404.

\bibitem{androulaki2013evaluating}
E.~Androulaki, G.~O. Karame, M.~Roeschlin, T.~Scherer, and S.~Capkun,
  ``Evaluating user privacy in bitcoin,'' in \emph{IFCA}.\hskip 1em plus 0.5em
  minus 0.4em\relax Springer, 2013, pp. 34--51.

\bibitem{bauer1972constructing}
D.~F. Bauer, ``Constructing confidence sets using rank statistics,''
  \emph{Journal of the American Statistical Association}, vol.~67, no. 339, pp.
  687--690, 1972.

\bibitem{Carlsson:2009}
G.~Carlsson, ``Topology and data,'' \emph{Bul. of the AMS}, vol.~46, no.~2,
  2009.

\bibitem{chen2016xgboost}
T.~Chen and C.~Guestrin, ``Xgboost: A scalable tree boosting system,'' in
  \emph{The 22nd {SIGKDD}}.\hskip 1em plus 0.5em minus 0.4em\relax ACM, 2016,
  pp. 785--794.

\bibitem{georgia-ransomware}
C.~Cimpanu, ``Georgia county pays a whopping \$400,000 to get rid of a
  ransomware infection
  \url{https://www.zdnet.com/article/georgia-county-pays-a-whopping-400000-to-get-rid-of-a-ransomware-infection/},''
  \emph{ZDNet}, Mar 2019.

\bibitem{conti2018economic}
M.~Conti, A.~Gangwal, and S.~Ruj, ``On the economic significance of ransomware
  campaigns: A bitcoin transactions perspective,'' \emph{Computers \&
  Security}, 2018.

\bibitem{di2015bitconeview}
G.~Di~Battista, V.~Di~Donato, M.~Patrignani, M.~Pizzonia, V.~Roselli, and
  R.~Tamassia, ``Bitconeview: visualization of flows in the bitcoin transaction
  graph,'' in \emph{IEEE VizSec}, 2015, pp. 1--8.

\bibitem{dingledine2004tor}
R.~Dingledine, N.~Mathewson, and P.~Syverson, ``Tor: The second-generation
  onion router,'' Naval Research Lab Washington DC, Tech. Rep., 2004.

\bibitem{ester1996density}
M.~Ester, H.-P. Kriegel, J.~Sander, X.~Xu \emph{et~al.}, ``A density-based
  algorithm for discovering clusters in large spatial databases with noise.''
  in \emph{Kdd}, vol.~96, no.~34, 1996, pp. 226--231.

\bibitem{fleder2015bitcoin}
M.~Fleder, M.~S. Kester, and S.~Pillai, ``Bitcoin transaction graph analysis,''
  \emph{arXiv preprint arXiv:1502.01657}, 2015.

\bibitem{Forgy65}
E.~Forgy, ``Cluster analysis of multivariate data: efficiency versus
  interpretability of classifications,'' \emph{Biometrics}, vol.~21, pp.
  768--780, 1965.

\bibitem{greaves2015using}
A.~Greaves and B.~Au, ``Using the bitcoin transaction graph to predict the
  price of bitcoin,'' \emph{No Data}, 2015.

\bibitem{surveykent}
J.~Hernandez-Castro, E.~Boiten, and M.~Barnoux, ``The 2nd kent cyber security
  survey,'' \emph{Kent University reports}, 2014,
  https://kar.kent.ac.uk/52891/.

\bibitem{tin1995RF}
T.~K. Ho, ``Random decision forests,'' in \emph{Proceedings of 3rd
  International Conference on Document Analysis and Recognition}, vol.~1, Aug
  1995, pp. 278--282 vol.1.

\bibitem{huang2018tracking}
D.~Y. Huang, D.~McCoy, M.~M. Aliapoulios, V.~G. Li, L.~Invernizzi,
  E.~Bursztein, K.~McRoberts, J.~Levin, K.~Levchenko, and A.~C. Snoeren,
  ``Tracking ransomware end-to-end,'' in \emph{Tracking Ransomware
  End-to-end}.\hskip 1em plus 0.5em minus 0.4em\relax IEEE, 2018, pp. 1--12.

\bibitem{APWG}
J.~Lewis, ``Economic impact of cybercrime -- no slowing down,'' \emph{McAfee
  Report}, no.~2, 2018.

\bibitem{liao2016behind}
K.~Liao, Z.~Zhao, A.~Doup{\'e}, and G.-J. Ahn, ``Behind closed doors:
  measurement and analysis of cryptolocker ransoms in bitcoin,'' in \emph{2016
  APWG Symposium on Electronic Crime Research (eCrime)}.\hskip 1em plus 0.5em
  minus 0.4em\relax IEEE, 2016, pp. 1--13.

\bibitem{lilliefors1967kolmogorov}
H.~W. Lilliefors, ``On the kolmogorov-smirnov test for normality with mean and
  variance unknown,'' \emph{Journal of the American statistical Association},
  vol.~62, no. 318, pp. 399--402, 1967.

\bibitem{lum2013extracting}
P.~Y. Lum, G.~Singh, A.~Lehman, T.~Ishkanov, M.~Vejdemo-Johansson,
  M.~Alagappan, J.~Carlsson, and G.~Carlsson, ``Extracting insights from the
  shape of complex data using topology,'' \emph{Scientific reports}, vol.~3, p.
  1236, 2013.

\bibitem{maaten2008visualizing}
L.~v.~d. Maaten and G.~Hinton, ``Visualizing data using t-sne,'' \emph{Journal
  of machine learning research}, vol.~9, no. Nov, pp. 2579--2605, 2008.

\bibitem{martin2018depth}
A.~Martin, J.~Hernandez-Castro, and D.~Camacho, ``An in-depth study of the
  jisut family of android ransomware,'' \emph{IEEE Access}, vol.~6, pp.
  57\,205--57\,218, 2018.

\bibitem{maxwell2013coinjoin}
G.~Maxwell, ``Coinjoin: Bitcoin privacy for the real world,'' in \emph{Post on
  Bitcoin Forum}, 2013.

\bibitem{mcginn2016visualizing}
D.~McGinn, D.and~Birch, D.~Akroyd, M.~Molina-Solana, Y.~Guo, and W.~J.
  Knottenbelt, ``Visualizing dynamic bitcoin transaction patterns,'' \emph{Big
  data}, vol.~4, no.~2, pp. 109--119, 2016.

\bibitem{meiklejohn2013fistful}
S.~Meiklejohn, M.~Pomarole, G.~Jordan, D.~Levchenko, K.and~McCoy, G.~M.
  Voelker, and S.~Savage, ``A fistful of bitcoins: characterizing payments
  among men with no names,'' in \emph{Proceedings of the 2013 conference on
  Internet measurement conference}.\hskip 1em plus 0.5em minus 0.4em\relax ACM,
  2013, pp. 127--140.

\bibitem{nakamoto2008bitcoin}
S.~Nakamoto, ``Bitcoin: A peer-to-peer electronic cash system,'' 2008.

\bibitem{narayanan2017obfuscation}
A.~Narayanan and M.~M{\"o}ser, ``Obfuscation in bitcoin: Techniques and
  politics,'' \emph{arXiv preprint arXiv:1706.05432}, 2017.

\bibitem{ober2013structure}
M.~Ober, S.~Katzenbeisser, and K.~Hamacher, ``Structure and anonymity of the
  bitcoin transaction graph,'' \emph{Future internet}, vol.~5, no.~2, pp.
  237--250, 2013.

\bibitem{paquet2018ransomware}
M.~Paquet-Clouston, B.~Haslhofer, and B.~Dupont, ``Ransomware payments in the
  bitcoin ecosystem,'' \emph{arXiv preprint arXiv:1804.04080}, 2018.

\bibitem{WSJ2018}
C.~Ramey, ``The crypto crime wave is here,'' \emph{Wall Street Journal}, no.
  April 28, 2018.

\bibitem{ron2013quantitative}
D.~Ron and A.~Shamir, ``Quantitative analysis of the full bitcoin transaction
  graph,'' in \emph{International Conference on Financial Cryptography and Data
  Security}.\hskip 1em plus 0.5em minus 0.4em\relax Springer, 2013, pp. 6--24.

\bibitem{ruffing2014coinshuffle}
T.~Ruffing, P.~Moreno-Sanchez, and A.~Kate, ``Coinshuffle: Practical
  decentralized coin mixing for bitcoin,'' in \emph{European Symposium on
  Research in Computer Security}.\hskip 1em plus 0.5em minus 0.4em\relax
  Springer, 2014, pp. 345--364.

\bibitem{scaife2016cryptolock}
N.~Scaife, H.~Carter, P.~Traynor, and K.~R. Butler, ``Cryptolock (and drop it):
  stopping ransomware attacks on user data,'' in \emph{2016 IEEE 36th
  International Conference on Distributed Computing Systems (ICDCS)}.\hskip 1em
  plus 0.5em minus 0.4em\relax IEEE, 2016, pp. 303--312.

\bibitem{singh2007topological}
G.~Singh, F.~M{\'e}moli, and G.~E. Carlsson, ``Topological methods for the
  analysis of high dimensional data sets and 3d object recognition.'' in
  \emph{SPBG}, 2007, pp. 91--100.

\bibitem{spagnuolo2014bitiodine}
M.~Spagnuolo, F.~Maggi, and S.~Zanero, ``Bitiodine: Extracting intelligence
  from the bitcoin network,'' in \emph{International Conference on Financial
  Cryptography and Data Security}.\hskip 1em plus 0.5em minus 0.4em\relax
  Springer, 2014, pp. 457--468.

\bibitem{swan2015blockchain}
M.~Swan, \emph{Blockchain: Blueprint for a new economy}.\hskip 1em plus 0.5em
  minus 0.4em\relax O'Reilly Media, Inc., 2015.

\bibitem{symantecwall}
Symantec, ``Ransom.cryptowall,'' \emph{Online}, 2019,
  \url{{www.symantec.com/security-center/writeup/2016-041912-5637-99}}.

\bibitem{tschorsch2016bitcoin}
F.~Tschorsch and B.~Scheuermann, ``Bitcoin and beyond: A technical survey on
  decentralized digital currencies,'' \emph{IEEE Communications Surveys \&
  Tutorials}, vol.~18, no.~3, pp. 2084--2123, 2016.

\end{thebibliography}

\end{document}